\documentclass[fleqn,usenatbib]{mnras}

\usepackage{newtxtext,newtxmath}

\usepackage[T1]{fontenc}

\DeclareRobustCommand{\VAN}[3]{#2}
\let\VANthebibliography\thebibliography
\def\thebibliography{\DeclareRobustCommand{\VAN}[3]{##3}\VANthebibliography}

\usepackage{graphicx}	
\usepackage{amsmath}	
\usepackage{multirow}
\usepackage{adjustbox}
\DeclareUnicodeCharacter{2212}{-}

\title[NGC\,4666 multi-phase galactic-scale wind]{The GECKOS Survey: Resolved, multiphase observations of mass-loading and gas density in the galactic wind of NGC\,4666}

\author[B. Mazzilli Ciraulo et al.]{B. Mazzilli Ciraulo$^{1,2}$\thanks{E-mail: bmazzilliciraulo@swin.edu.au}, D.~B. Fisher$^{1,2}$, R. Elliott$^{1,2}$, A. Fraser-McKelvie$^{3,2}$, M.~R. Hayden$^4$, M. Martig$^5$,
\newauthor J. van de Sande$^{6,2}$, A. J. Battisti$^{7,8,2}$, J. Bland-Hawthorn$^{9,2}$, A. D. Bolatto$^{10}$, T. H. Brown$^{11}$, B. Catinella$^{7,2}$,
\newauthor F. Combes$^{12,13}$, L. Cortese$^{7,2}$, T. A. Davis$^{14}$, E. Emsellem$^3$, D. A. Gadotti$^{15}$, C. del P. Lagos$^{7,2,16}$, X. Lin$^{17}$,
\newauthor A. Marasco$^{18}$, E. Peng$^{19}$, F. Pinna$^{20,21}$, T. H. Puzia$^{22}$, L. A. Silva-Lima$^{23}$, L. M. Valenzuela$^{24}$,
\newauthor G. van de Ven$^{25}$, J. Wang$^{17}$
\\
$^{1}$Centre for Astrophysics and Supercomputing, Swinburne University of Technology, Hawthorn, VIC 3122, Australia\\
$^{2}$ARC Centre of Excellence for All Sky Astrophysics in 3 Dimensions (ASTRO 3D), Australia\\
$^{3}$ European Southern Observatory, Karl-Schwarzschild-Stra{\ss}e 2, Garching, 85748, Germany \\
$^{4}$ Homer L. Dodge Department of Physics \& Astronomy, University of Oklahoma, 440 W. Brooks St., Norman, OK 73019, USA \\ 
$^{5}$ Astrophysics Research Institute, Liverpool John Moores University, 146 Brownlow Hill, Liverpool L3 5RF, UK \\
$^{6}$School of Physics, University of New South Wales, Sydney, NSW 2052, Australia \\
$^{7}$International Centre for Radio Astronomy Research (ICRAR), The University of Western Australia, M468, 35 Stirling Highway, Crawley, WA 6009, Australia \\
$^{8}$Research School of Astronomy and Astrophysics, Australian National University, Cotter Road, Weston Creek, ACT 2611, Australia \\
$^{9}$Sydney Institute for Astronomy, School of Physics, A28, The University of Sydney, NSW, 2006, Australia \\
$^{10}$Department of Astronomy, University of Maryland, College Park, MD 20742, USA \\
$^{11}$National Research Council of Canada, Herzberg Astronomy and Astrophysics Research Centre, 5071 W. Saanich Rd. Victoria, BC, V9E 2E7, Canada\\
$^{12}$Observatoire de Paris, LUX, CNRS, PSL University, Sorbonne University, 75014 Paris, France \\
$^{13}$ Coll\`ege de France, 11 Pl. Marcelin Berthelot, 75231 Paris, France \\
$^{14}$Cardiff Hub for Astrophysics Research \&\ Technology, School of Physics \&\ Astronomy, Cardiff University, Queens Buildings, Cardiff, CF24 3AA, UK \\
$^{15}$ Centre for Extragalactic Astronomy, Department of Physics, Durham University, South Road, Durham DH1 3LE, UK \\
$^{16}$ Cosmic Dawn Center (DAWN), Denmark \\
$^{17}$Kavli Institute for Astronomy and Astrophysics, Peking University, Beijing 100871, People's Republic of China \\
$^{18}$INAF – Padova Astronomical Observatory, Vicolo dell’Osservatorio 5, I-35122 Padova, Italy \\
$^{19}$ NSF's NOIRLab, 950 N. Cherry Avenue, Tucson, AZ 85719, USA \\
$^{20}$Instituto de Astrof\'isica de Canarias, calle Vía L\'actea s/n, E-38205 La Laguna, Tenerife, Spain\\
$^{21}$Departamento de Astrof\'isica, Universidad de La Laguna, Avenida Astrof\'isico Francisco S\'anchez s/n, E-38206 La Laguna, Spain \\
$^{22}$Instituto de Astrofísica, Pontificia Universidad Católica de Chile, Avenida Vicuña Mackenna 4860, 7820436, Macul, Santiago, Chile \\
$^{23}$Núcleo de Astrofísica, Universidade Cidade de São Paulo, Rua Galvão Bueno, 868, São Paulo, Brazil \\
$^{24}$Universitäts-Sternwarte, Fakultät für Physik, Ludwig-Maximilians-Universität München, Scheinerstr. 1, 81679 München,
Germany \\
$^{25}$ Department of Astrophysics, University of Vienna, T\"urkenschanzstra{\ss}e 17, 1180 Vienna, Austria \\
}

\date{Accepted XXX. Received YYY; in original form ZZZ}

\pubyear{2025}

\begin{document}
\label{firstpage}
\pagerange{\pageref{firstpage}--\pageref{lastpage}}
\maketitle

\begin{abstract}
We present a multiphase, resolved study of the galactic wind extending from the nearby starburst galaxy NGC\,4666. For this we use VLT/MUSE observations from the GECKOS program and \ion{H}{I} data from the WALLABY survey. We identify both ionised and \ion{H}{I} gas in a biconical structure extending to at least $z\sim$8~kpc from the galaxy disk, with increasing velocity offsets above the midplane in both phases, consistent with a multiphase wind.
The measured electron density, using [\ion{S}{II}],  differs significantly from standard expectations of galactic winds. We find electron density declines from the galaxy centre to $\sim2$~kpc, then rises again, remaining high ($\sim100-300$~cm$^{-3}$) out to $\sim$5~kpc. We find that \ion{H}{I} dominates the mass loading.
The total \ion{H}{I} mass outflow rate (above $z~>2$~kpc) is
between $5-13~M_{\odot}~\rm yr^{-1}$, accounting for uncertainties from disk-blurring and group interactions. The total ionised mass outflow rate (traced by H$\alpha$) is between $0.5~M_{\odot}~\rm yr^{-1}$ and $5~M_{\odot}~\rm yr^{-1}$, depending on $n_e(z)$ assumptions.
From ALMA/ACA observations, we place an upper-limit on CO flux in the outflow which correlates to $\lesssim2.9~M_{\odot}~\rm yr^{-1}$.
We also show that the entire outflow is not limited to the bicone, but a secondary starburst at the edge generates a more widespread outflow, which should be included in simulations. The cool gas in NGC\,4666 wind has insufficient velocity to escape the halo of a galaxy of its mass, especially because most of the mass is present in the slower atomic phase. This strong biconical wind contributes to gas cycling around the galaxy.
\end{abstract}




\section{Introduction}
\label{sec:intro}
Galaxy-scale winds, also referred to as outflows, are ubiquitous in high star formation rate (SFR) galaxies at all redshifts \citep[e.g.][]{Veilleux2005, Rubin2014, Davies2024, Carniani2024}. Simulations find that they play a dominant role in regulating galaxy evolution \citep[e.g.][]{NaabOstriker2017, Pillepich2018,ThompsonHeckman2024, Wright2024}. Winds are needed to reproduce many fundamental observable quantities, such as the stellar mass function and star formation rates of galaxies \citep[e.g.][]{SomervilleDave2015}. They contribute significantly to galaxy evolution by regulating star formation.
Even if the gas does not escape the halo, gas expelled from the midplane is unlikely to be readily available to form stars in the disk. In starburst environments observations indicate that more gas is removed in the outflow than converted to stars \citep{Bolatto2013b,Leroy2015, Fluetsch2019, ReichardtChu2022b}, making the wind the primary regulator of star formation.

Observations of the nearest galaxies hosting star-formation driven galactic winds \citep{Bolatto2013b,Leroy2015,Salak2020} show that cold gas (molecular and atomic) is the dominant mass component of galactic winds \citep[see][for a review]{Veilleux2020}. Recent \textit{JWST} observations of the M\,82 wind suggest that, at least near the disk, cold gas may be able to survive disk breakout in the form of filamentary structures and small-scale clouds \citep{Bolatto2024,Fisher2025}. In this scenario, cold gas lifted from the disk directly removes the fuel for star formation.  Moreover, theory and simulations argue that the interaction of warm-and-cold gas in outflows is a critical component to setting the velocities, and hence the ability of outflows to regulate star formation (\citealt{FieldingBryan2022,Nikolis2024}; for review see \citealt{ThompsonHeckman2024}). Yet, the rarity of observations of the coldest phases of gas, let alone multiphase observations, in galactic winds hampers the ability of observations to uniquely test theory. 

A picture has emerged in which high-velocity, ionised gas typically forms biconical structures above the plane of starburst galaxies \citep{ShopbellBlandHawthorn1998,VeilleuxRupke2002,Bik2018,Rupke2019,McPherson2023,Herenz2023,Watts2024}. Opening angles of those winds in these targets typically range between $20-30\degr$, which suggests a collimated flow of gas leaving a disk. In most cases the gas is detected to distances of 5-15~kpc. However, this extent is likely limited by observations, such as field-of-view and sensitivity. In very high SFR systems, however, winds of ionised gas extend to $\sim$50~kpc \citep{Rupke2019}. Observations of face-on galaxies, with similar mass and SFR as edge-on galaxies with biconical winds, determine gas velocity much more robustly and find ionised gas outflow velocities of order $\sim100-500$~km~s$^{-1}$ \citep{Newman2012, Davies2019, Avery2021, ReichardtChu2022a,ReichardtChu2024}, which is consistent with gas that is travelling tens of kiloparsecs away from the galaxy on timescales of $\sim10-100$ Myr. The mass loading factor of ionised gas (outflow mass rate divided by SFR of galaxy) is typically $\eta = \dot{M}_{\rm out}/\rm SFR\sim 0.1-3$ across a wide range of nearby galaxies \citep{McQuinn2019, Davies2019, Concas2019, ReichardtChu2022a, Yuan2023, McPherson2023, Watts2024}. 

There is a considerable amount of uncertainty in estimating ionised gas mass loading \citep[discussed in][]{Yuan2023}. 
This is because key quantities required for such estimates — electron density, spatial extent, and covering fraction — are rarely known for individual galaxies in surveys of galactic winds.
Very few resolved observations of electron density have been made in the literature, largely due to the observational challenge of detecting the faint emission lines required for density diagnostics. Yet electron density measurements are both important for estimating the mass loss in ionised gas \citep[described in][]{Veilleux2020}, but also as the electron density profile is a direct observable test of simulations and theory describing galactic winds \citep{ChevalierClegg1985,Schneider2020,FieldingBryan2022}. Classic adiabatic wind models require that the electron density follows a steeply declining power-law profile; which is observed in metal-poor outflows \citep{Bik2018,HamelBravo2024}.
Recent work by \cite{Xu2023}, however, finds that the electron density profile in M\,82 is more shallow than expectations from adiabatic winds.

Resolved observations of cold gas in winds are limited to a few nearby starburst galaxies \citep{Walter2002, Sakamoto2006, Tsai2012, Bolatto2013b, Kreckel2014, Martini2018, Salak2020,Bolatto2021}, but underline their importance as the cold phase of gas is often carrying most of the gas mass. Estimation of the total mass-loading and energy-loading of galactic winds, therefore, requires the observation of the cold phase of gas \citep{JBH2007}. 
Observations of M\,82 suggest that while the mass loading is dominated by colder phase (molecular and atomic), the energy rate carried by the ions in the wind is higher (\citealt{Leroy2015,Martini2018}; also see \citealt{Yuan2023,Xu2023}).
These differences in phases of the winds can provide a more complete picture to test those theories of outflow kinematics that are built on interactions of gas in different phases \citep{FieldingBryan2022}, and more multiphase observations are needed. The advent of integral field spectrographs on 8\,m telescopes has allowed for extremely sensitive observations of faint gas, recovering faint emission features in the optical regime. Observations using ALMA have provided a view of the nearest winds over the past decade \citep[see review by][]{Veilleux2020}. The impending commissioning of the Square Kilometer Array (SKA), toward the end of this decade, presents the opportunity to expand the sample. Currently, data from SKA-precursors, MeerKAT and Australian Square Kilometre Array Pathfinder (ASKAP) allow for some increase in the capacity to observe the cold phase of winds. In this paper we will focus on multiphase observations of the galactic wind in NGC\,4666.

NGC\,4666 is a highly-inclined ($\sim70\degr$, HyperLeda\footnote{http://leda.univ-lyon1.fr/}) spiral galaxy at a distance of 15.5~Mpc \citep{Tully2023}. NGC\,4666 is a Milky Way mass galaxy, $M_{\star}\sim 7\times 10^{10}$~$M_{\odot}$ \citep{Sheth2010}. The star formation rate is SFR$\sim$10~M$_{\odot}$~yr$^{-1}$ \citep{Vargas2019}, which is 6$\times$ higher than the main-sequence value for its mass \citep{Popesso2023}. It is a gas-rich system, with $M_{\rm \ion{H}{I}}/M_{\star}\sim0.05$ \citep{Lee2022}, which is 3$\times$ higher than local Universe galaxies of a similar mass \citep{Catinella2018}.
\citet{Lee2022} discussed the CO(1-0) distribution of NGC\,4666 based on Atacama Compact Array observations and derived a high molecular surface density in the disk. They also detected CN, a tracer of dense molecular gas, in the central region.
Consistent with expectations of being gas-rich and starbursting, NGC\,4666 is known to host a galactic scale wind, highlighted by an outflow cone detected in optical imaging and spectroscopy \citep{Dahlem1997} and X-ray observations \citep{Tullmann2006}. The diffuse ionised gas in the halo shows an ``X-shaped'' structure and most of the X-ray emitting extraplanar gas appears to be confined in this filamentary Diffuse Interstellar Gas (DIG) structure \citep{Tullmann2006}. \cite{Lu2023} find in a sample of 22 nearby edge-on galaxies studied with narrow-band imaging that NGC\,4666 has the largest H$\alpha$ scale-height ($2.492\pm0.023$~kpc) along the vertical axis, which is consistent with gas being removed via an outflow.
\citet{Walter2004} also analysed high-resolution CO(1-0) observations and used the position-velocity (PV) diagram to interpret a kinematic feature as an expanding molecular shell. This super-shell is observed close to the location where the most prominent H$\alpha$ filament seems to emerge from.

Recently, \citet{Stein2019} found a point source and bubble structure in 6\,GHz and 1.5\,GHz radio continuum imaging. They argue that this observation suggests an Active Galactic Nucleus (AGN). \citet{Lee2022} detects extended 3\,mm continuum emission, interpreted as a signature of nuclear activity in the galactic centre, potentially from an obscured AGN.
However, in optical imaging there is no evidence of a strong nuclear point source nor AGN-like emission line ratios \citep{LopezCoba2020}. Moreover, \cite{Persic2004} do not find a nuclear X-ray source in high-resolution imaging, which favours the absence of an AGN that is energetically contributing to the galaxy.
If there is some AGN activity, it does not seem to be currently energetically significant compared to the star formation. We note that a comparison of the wind energetics to the energy from star formation will be informative for assessing whether star formation alone can drive the wind in NGC\,4666.

NGC\,4666 is part of a group of galaxies and is interacting with the less massive NGC\,4668 and a dwarf companion, as revealed by deep \ion{H}{I} observations from the Very Large Array \citep[VLA;][]{Walter2004}. Recent \ion{H}{I} observations as part of the Widefield ASKAP $L$-band Legacy All-sky Blind surveY (WALLABY, \citealt{Koribalski2020}) pilot survey support this conclusion by exhibiting a ``peculiar'' \ion{H}{I} distribution, interpreted as the signature of the interaction occurring with the neighbouring galaxy NGC\,4668 \citep{Lee2022}.
This interaction is posited to have triggered the starburst activity in NGC\,4666, which in turn launched the superwind. This scenario is consistent with our understanding that starbursts in the Local Universe are initiated by galaxy mergers or interactions \citep[e.g.][]{Bournaud2011,Renaud2022}.
Table~\ref{tab:ngc4666} summarises the properties of this galaxy reported in the literature.

While previous studies have established the presence of a galactic wind, many open questions remain about the resolved structure, kinematics, properties of the extraplanar gas. To address this, we use observations from the Multi Unit Spectroscopic Explorer (MUSE) on the Very Large Telescope (VLT), which provide integral-field spectroscopy with high spatial resolution and full spectral coverage across the optical range. This allows us to map the wind extent, analyse the ionised gas kinematics, and study the vertical extent of the electron density in the wind. These capabilities are essential to constrain the driving mechanism of the wind and to understand the cycling of gas surrounding NGC\,4666.

We present VLT/MUSE and the ancillary observations of NGC\,4666 in Sect.~\ref{sec:obs_reduction}. Section~\ref{sec:wind_def_sigma_gas} describes the superwind structure as well as the star formation properties in the disk underneath it. We present the wind kinematic sub-structure in Sect.~\ref{sec:wind_kinematics}. We discuss the electron density profile inferred from VLT/MUSE observations and its implications for outflow rate estimates in Sect.~\ref{sec:ne_profile}. Section~\ref{sec:outflow_rates} presents the outflow mass and energy loading determinations. Section~\ref{sec:bubble} focuses on the properties of extraplanar emission outside of the bicone. We summarise and discuss our findings in Sect.~\ref{sec:summary}.

\begin{table*}
    \caption{Properties of NGC\,4666.}
    \label{tab:ngc4666}
    \centering
    \begin{tabular}{lccc}
        \hline
        Property & Value & Unit & Reference \\
        \hline
        R.A. (J2000) & 12:45:08.591 & hms & \\
        Dec. (J2000) & $-$00:27:42.79 & dms &\\ 
        Distance & 15.5 & Mpc & \citet{Tully2023} \\
        Inclination & 69.6 & deg & HyperLEDA$^*$  \\
        Position angle & 40.6 & deg & HyperLEDA \\
        Stellar mass & 10.992 & log($M_{\star} / M_{\odot}$) & $3.6\micron+4.5\micron$, \citet{Sheth2010}$^{\rm a}$ \\
        Star formation rate & 7.29$\pm$1.82 & $M_{\odot}\cdot\,\rm yr^{-1}$ & $22\micron$, \citet{Wiegert2015} \\
            & 10.5$\pm$1.92 & &  H$\alpha$+$22\micron$, \citet{Vargas2019} \\
        $\frac{S_{\rm 60\micron}}{S_{\rm 100\micron}}$ & 0.4318 & & \citet{Sanders2003} \\
        $F_{\rm H\alpha}$ & $4.20\pm0.69$ & $\rm 10^{-12}~erg~s^{-1}~cm^{-2}$ & \citet{Vargas2019} \\
        Molecular gas mass & $9.46\pm0.02^{\rm b}$ & log($M_{\rm H_2} / M_{\odot}$)& \citet{Lee2022} \\
            & $9.40\pm0.02^{\rm c}$ & & \citet{Lee2022} \\
        \ion{H}{I} gas mass & 9.7 & log($M_{\rm \ion{H}{I}} / M_{\odot}$) & \citet{Westmeier2022}$^{\rm d}$  \\
        \hline
    \end{tabular}
    \parbox{0.8\linewidth}{\small \textit{Note.}
        $^*$ The HyperLEDA database is accessible via this link: http://leda.univ-lyon1.fr/.\\
        $^{\rm a}$ The stellar gas mass estimate is given for $D_L=15.5~\rm Mpc$ instead of their distance measurement.
        The 60-to-100 $\micron$ colour serves as proxy for dust temperature in regions of active star formation. A galaxy with a value typically exceeding $\gtrsim$0.4 is commonly identified as a starburst.\\
        The molecular gas mass estimates are derived from the CO luminosity using a Milky-Way like CO-to-H$_2$ conversion factor ($^{\rm b}$) and a metallicity-dependent CO-to-H$_2$ conversion factor ($^{\rm c}$).\\ $^{\rm d}$ The \ion{H}{I} gas mass estimate is given for $D_L=15.5~\rm Mpc$ instead of their distance measurement.}
\end{table*}

\section{Observations, data reduction and methods}
\label{sec:obs_reduction}
Our analysis is based on deep VLT/MUSE observations of the ionised gas, as well as archival ALMA/ACA observations and ASKAP observations. We describe these data and the data reduction methods in this section.

NGC\,4666 is targeted as part of the GECKOS project. GECKOS (Generalising Edge-on galaxies and their Chemical bimodalities, Kinematics, and Outflows out to Solar environments)\footnote{\url{https://geckos-survey.org/}} is a VLT/MUSE Large Program (317 hours, P.I.: Jesse van de Sande) targeting 36 edge-on galaxies \citep{GECKOSpaper}. These targets are selected from the S$^4$G survey \citep[66\,\%,][]{Sheth2010} and HyperLEDA \citep[34\,\%,][]{Makarov2014} within a distance range of $10<D<70\,\rm Mpc$ and with a stellar mass within $\pm0.3\,\rm dex$ of the Milky Way \citep[$5\times10^{10}~M_{\odot}$,][]{BlandHawthorn_Gerhard_2016}. With stellar mass controlled for, the remaining drivers of galaxy evolution for Milky Way-mass galaxies may be examined. The GECKOS sample spans a 2\,dex range in SFR ($\sim0.01-15~M_{\odot}~yr^{-1}$, based on the WISE W4 band). The deep MUSE observations commenced in 2022 and aim at studying the outflows, vertical structure of the interstellar medium (ISM), assembly history of stellar components, and chemical enrichment of Milky Way mass galaxies in the nearby Universe.

\subsection{MUSE observations and reduction}
NGC\,4666 was observed with MUSE, an integral field spectrograph mounted on the Unit Telescope 4 of the VLT. Observations were performed in wide field mode and using the nominal wavelength range, providing a field of view of $\rm \sim 1~arcmin\times1~arcmin$, a spectral sampling of $1.25\rm \AA~pixel^{-1}$ over the 4800--9300\AA\ wavelength coverage, and a spatial scale of $0.2~\rm arcsec~pixel^{-1}$.

In total, our mosaic consists of five MUSE pointings (two from GECKOS and three archival). The footprint of the MUSE mosaic is overlaid in the top-right panel of Figure~\ref{fig:ha_bicone}. This coverage enables us to probe an important part of the galaxy disk, as well as a region up to 8\,kpc off the plane in the north-west direction. For the GECKOS program, the galaxy was observed with two MUSE pointings, one targeting the upper side right H$\alpha$ limb (referred to as `outflow pointing' hereafter) and one targeting the upper side ``bubble'' (annotated in the left panel of Fig.~\ref{fig:ha_bicone}, referred to as `disk pointing' hereafter), which were placed to have an overlap of $2\arcsec$ with existing pointings available in the ESO archive (program ID:096.D-0296, PI: Anderson). Two archival pointings are closer to the galaxy centre and the third one extends to $\rm \sim 5\,kpc$ on the opposite side of the disk compared to the GECKOS outflow pointing. 

Each GECKOS observing block (OB) consists of four object (O) exposures with an integration time of 512\,s each, and two sky (S) exposures with an integration time of 120\,s each, executed in the order OSOOSO. There were no offset sky frames taken for the archival observations and each observing block had 4 object exposures of 702\,s each.
We perform the MUSE data reduction using the dedicated Python package \texttt{pymusepipe}\footnote{\url{https://pypi.org/project/pymusepipe/}} \citep{Emsellem2022} (version 2.27.2) which essentially consists of a wrapper around the different recipes included in the MUSE data processing pipeline \citep{Weilbacher2016,Weilbacher2020} and the ESO Recipe Execution Tool \citep[\texttt{esorex};][]{esorex2015}.
Firstly, we combine and use the calibration files to remove the instrument signature. We apply a flux calibration using a standard star observed on the same night as the OB. Then, we check and correct the astrometry solution for each MUSE science exposure by aligning them using Legacy Survey r-band imaging \citep{Dey2019}. From these aligned object exposures, we create one data cube per pointing with its associated World Coordinate System. The final stage of \texttt{pymusepipe} consists of creating a mosaicked cube by combining these five pointing cubes into one final data cube.
The archival observations do not include any sky frame; to remove the sky continuum in these frames, we define sky regions by selecting emission-line free regions in the frames and passing them as a sky mask to the \texttt{muse\_scipost} recipe. Finally, we account for differences in fluxes between GECKOS and archival pointings.
These differences likely arise from variations in observing conditions (e.g. atmospheric transparency, seeing), instrumental calibration differences between observing runs, or uncertainties in the flux calibration process itself.
The GECKOS `disk pointing' covers the south-west edge of the galaxy disk as well as the extraplanar region, extending up to $\sim$4~kpc from the midplane (see Fig.\,\ref{fig:ha_bicone}).
We assume that this GECKOS `disk pointing' provides the most reliable flux calibration for two reasons: first, as part of our observing program, it includes dedicated offset sky frames which allow for more accurate sky subtraction compared to the archival data; second, it targets brighter emission than the `outflow pointing', making the flux calibration more robust during the data reduction process.
To correct for flux differences, we compare the shape of the continuum in the overlapping region between this pointing and the adjacent archival pointing. We fit a third-order polynomial and shift the spectra in all spaxels within the adjacent archival pointing by the corresponding polynomial values.
For example, the flux corrections applied range from approximately $2.8~10^{-20}\rm erg~s^{-1}~cm^{-2}~$\AA$^{-1}$ at 4800~\AA\, to $3.9~10^{-20}\rm erg~s^{-1}~cm^{-2}~$\AA$^{-1}$ at 7000~\AA\, between the GECKOS disk pointing and the adjacent archival pointing.
We reproduce this procedure to scale all pointings of the mosaic. 

\subsubsection{Continuum subtraction}
We perform stellar continuum subtraction using the \texttt{nGIST} pipeline (version 5.3.0) \footnote{\url{https://geckos-survey.github.io/gist-documentation/}} \citep{nGIST,FraserMcKelvie2025}, an extension of the Galaxy IFU Spectroscopy Tool (GIST) pipeline \citep{Bittner2019}, comprising significant science improvements and performance updates, including the possibility of creating continuum-subtracted and line-only cubes. Through its CONT module, \texttt{nGIST} employs the \texttt{pPXF} full spectral fitting routine \citep{CappellariEmsellem2004,Cappellari2017}. We bin all spaxels to signal-to-noise ratio $\rm S/N=7$ using the Voronoi adaptive spatial binning method and use differential stellar population models from \citet{Walcher2009}. We fit the binned spectra over the rest-frame wavelength range $4750-7100~$\AA~ using 13th-order multiplicative polynomials only. The Milky Way foreground extinction is included in the pipeline and uses the \citet{Cardelli1989} law. The best-fit continuum spectrum in each Voronoi bin is rescaled to match the flux level of each individual spaxel and then subtracted.

\subsubsection{Emission line fitting}
We create an emission-line data cube by removing the continuum modelled by \texttt{nGIST} from the raw MUSE cube. We then measure all emission line properties using this cube. We spatially bin the mosaicked emission-line cube in 3$\times$3 spaxels to have a spaxel size of 0.6\arcsec $\times$ 0.6\arcsec, which is a typical seeing value in Paranal. We then carry out emission line fitting using the specialised emission line fitting package \texttt{threadcount}\footnote{\url{https://threadcount.readthedocs.io}}. This is an upgraded version of the one described in \cite{McPherson2023}. The software uses a bespoke version of the Python package \texttt{lmfit}\footnote{https://pypi.org/project/lmfit/}, which is optimised for speed, and employs the \texttt{nelder} minimisation algorithm. We fit H$\beta$, [\ion{O}{III}]~$\lambda$5007, [\ion{N}{II}]~$\lambda$6548, H$\alpha$, [\ion{N}{II}]~$\lambda$6583, [\ion{S}{II}]~$\lambda$6716 and [\ion{S}{II}]~$\lambda$6731. GECKOS observations feature an extremely large range in surface brightness per spaxel (from the bright disk centre to the extraplanar regions), and therefore \texttt{threadcount} is designed with this in mind. A rough $S/N$ for each emission line in each spaxel is calculated by direct integration of the flux and comparison to the variance extension of the MUSE cube.
We carry out fitting with two separate settings depending on the estimated emission line $S/N$, in order to obtain robust measurements for noisier spaxels: For spaxels with $S/N>30$, we run a single fit to the spectral line. The boundary of $S/N=30$ is chosen based on the runtime of the software. For all spaxels with $S/N<30$ we run 10 separate fits to the spectral line. The software creates simulated spectra by modifying the observed flux in each spectral pixel, based on a normal distribution whose standard deviation is determined by the observed variance in the data. The fit parameters and uncertainty are then taken as the mean and standard deviation of those fits. We fit each emission line using a single Gaussian model.

The internal dust attenuation was then determined for each spaxel using the Balmer decrement, H$\alpha$/H$\beta$, and corrected using a \citet{Cardelli1989} extinction law, with $R_v=3.1$. We assume an intrinsic Balmer decrement H$\alpha$/H$\beta=2.87$, which is typical for gas at a temperature of about $10^4~$K \citep{OsterbrockFerland2006}. For those spaxels without sufficient flux ($S/N>5$) in H$\beta$ we interpolate the extinction from the surrounding region. 
We find a median extinction $A_V\approx1.4$ and $A_{\rm H\alpha}\approx1.1$, which are consistent with the values reported by \citet{Voigtlander2013} for this galaxy: $A_V=1.61$ and $A_{\rm H\alpha}=1.15$. Within the inner kiloparsec parallel to the major axis of NGC\,4666, the median values we find become $A_V\approx1.9$ and $A_{\rm H\alpha}\approx1.6$.

\subsection{Narrow-band imaging data}
\label{sec:ha_imaging}
The H$\alpha$ image is derived from deep \(g\), \(r\) and \(N662\)-band images obtained at the CTIO/DECam, as part of a wide-field (30 deg$^2$) observing program of the NGC\,4636 group (PI: E. Peng). The DECam Community Pipeline is exploited to obtain calibrated images. Then a customised Python pipeline is used to build source masks, subtract backgrounds, and stack and co-add images into science images of the three bands. We combine \(g\) and \(r\) band images, following the procedure described in \cite{Boselli2015}, to derive the image of continuum fluxes underlying the \(N662\) narrow band. Then the continuum images are subtracted from the \(N662\)-band images, and the H$\alpha$ images are obtained. The spatial sampling is 0.27~arcsec~pixel$^{-1}$.

\subsection{ALMA/ACA observations}
We use CO(1-0) observations performed with the Atacama Compact Array (ACA), available in the ALMA science archive (project ID: 2019.1.01804.S; PI: B. Lee). The observations were tuned to a sky frequency of 114.70~GHz. Observations were calibrated and cleaned using the Common Astronomy Software Applications package (CASA, version 5.6.1-8). The maximum recoverable scale is $\sim78^{\prime\prime}$, which corresponds to $\sim5.9$~kpc. The final data cube, binned into a velocity resolution of $\rm 20~km~s^{-1}$, has a pixel size of $2\arcsec$ and a synthesised beam size of $\rm \sim13.9\times9.6\,arcsec^2$ ($\rm \sim1.05\,kpc\times722\,pc$). The rms noise level per channel is $\rm 11.1\,mJy~beam^{-1}$. For further details on the observations and data reduction, see \citet{Lee2022}.

\begin{figure*}
    \centering
    \includegraphics[width=0.99\columnwidth]{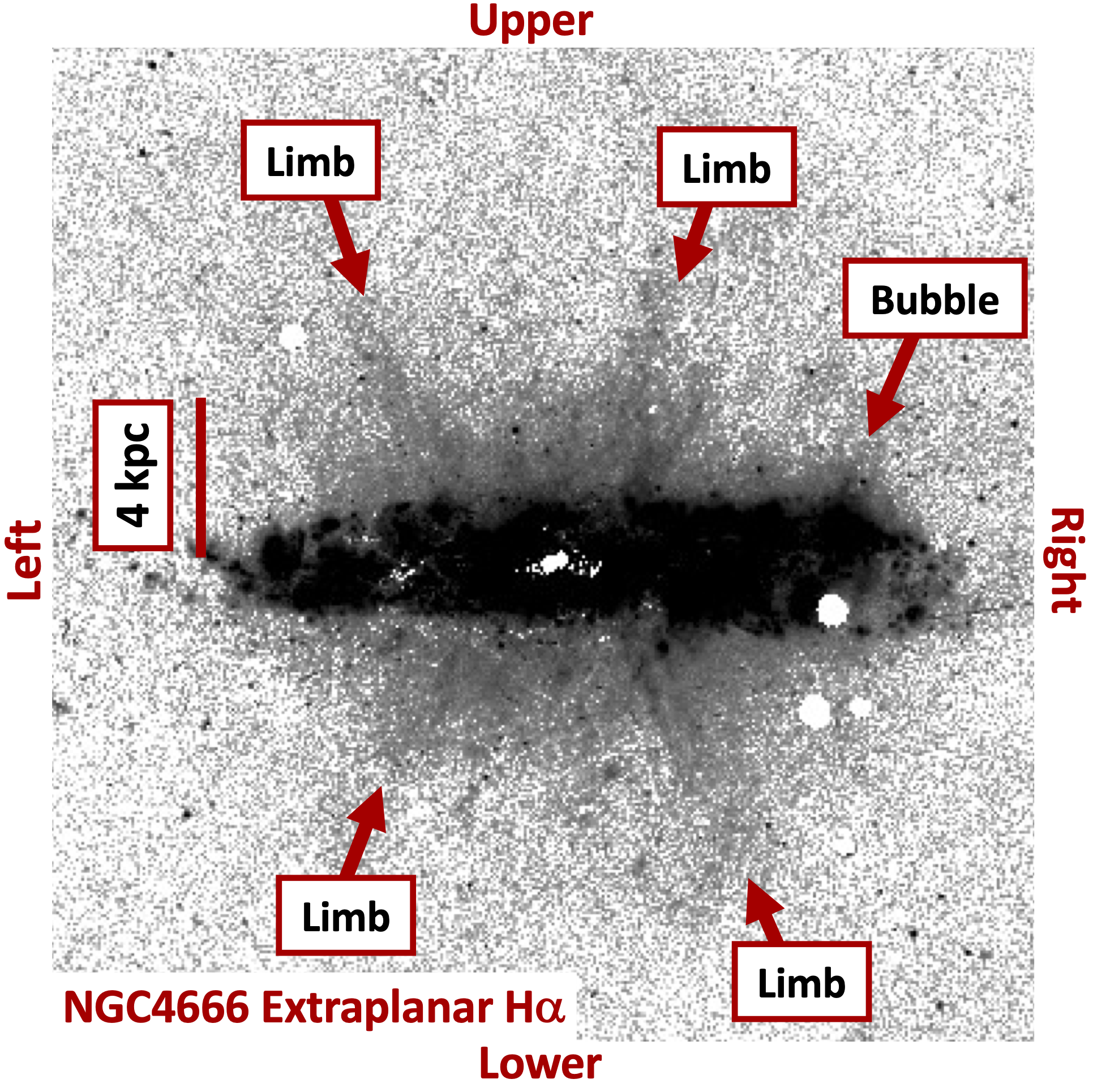}
    \includegraphics[width=1.01\columnwidth]{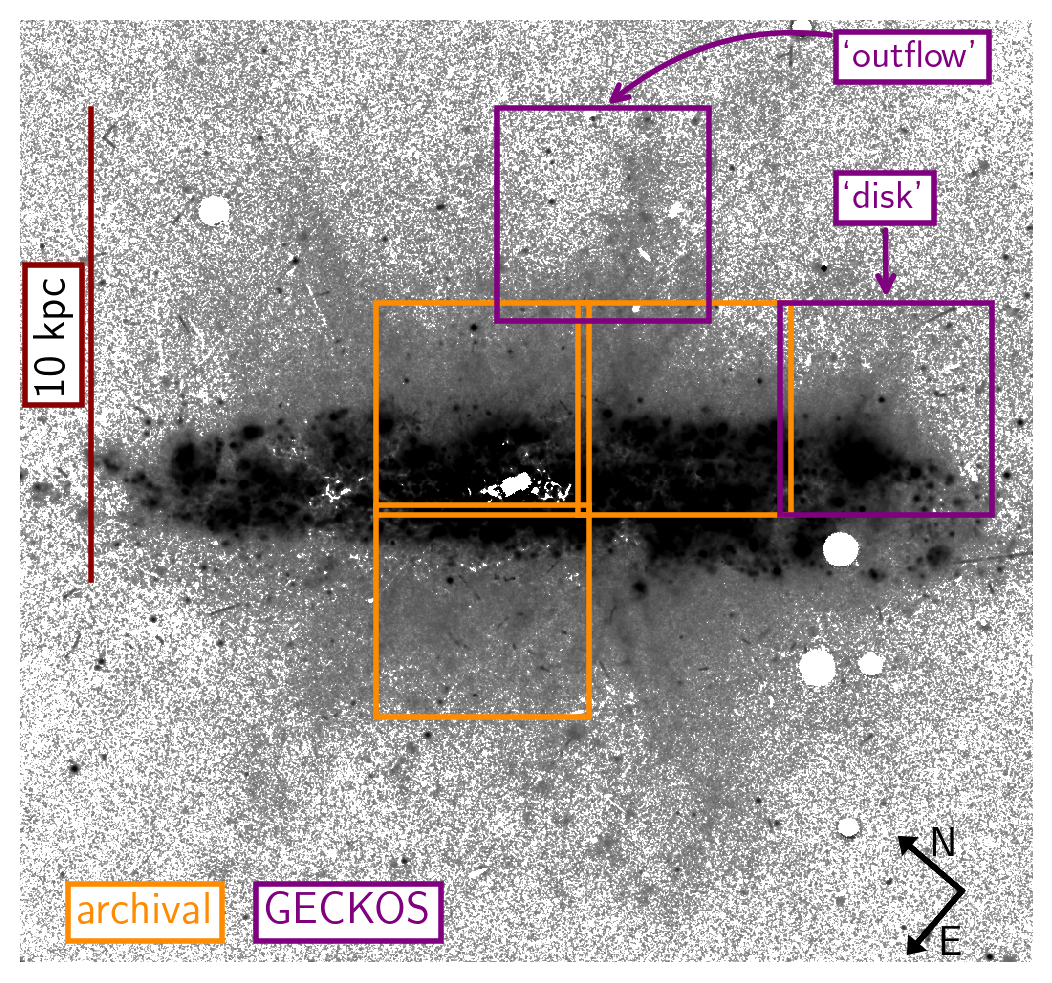}\\
    \includegraphics[width=0.7\textwidth]{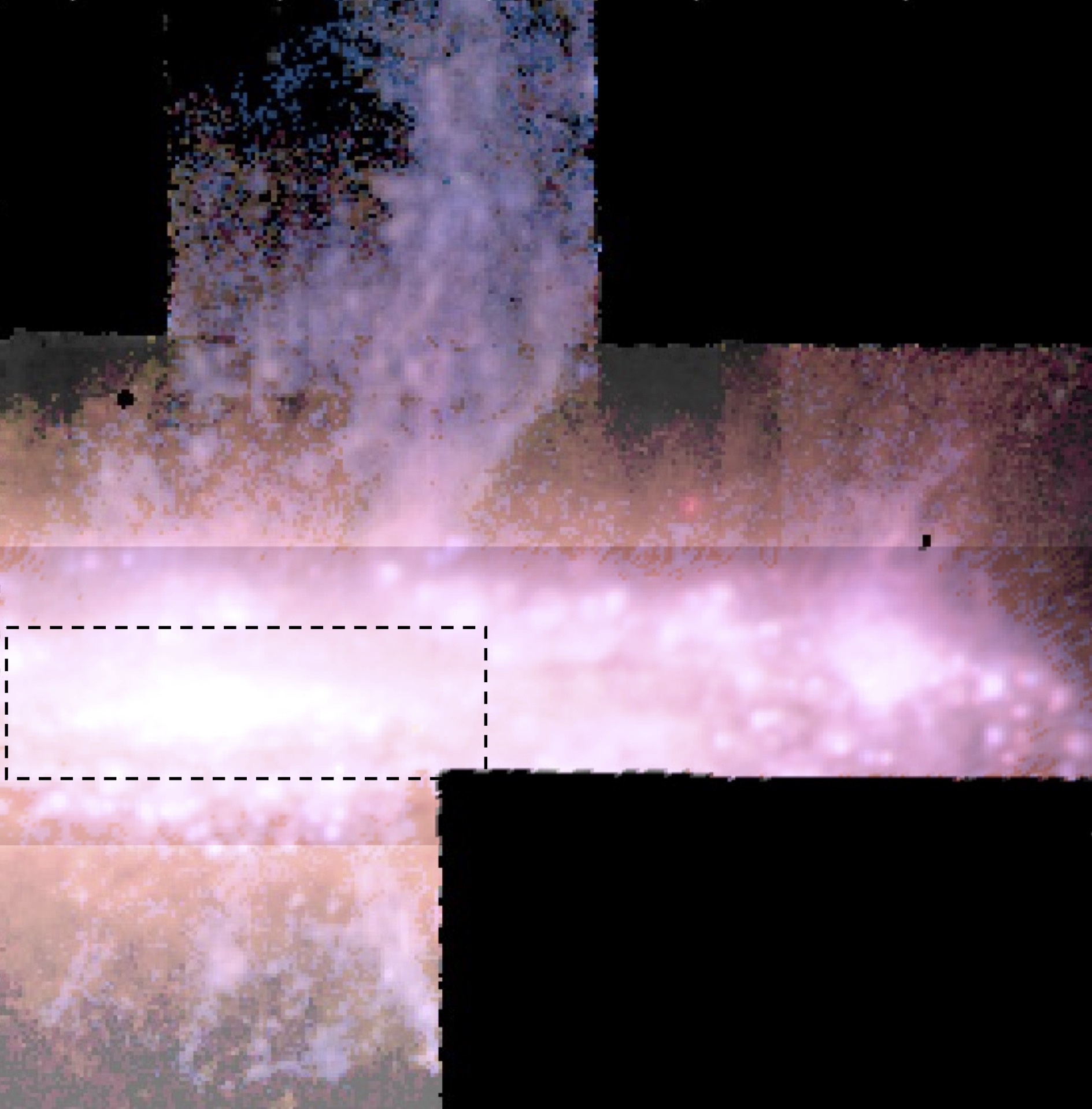}

    \caption{ {\bf Top left:} H$\alpha$ narrow-band image of NGC\,4666, rotated such that the disk is horizontal. The image is plotted in log-scale to highlight the low surface brightness components. The biconical superwind is denoted by the red arrows. We identify 4 filaments, called `limbs', that form an X-shape typical of a biconical wind. The H$\alpha$ gas is brighter inside the limbs than outside. We also identify a smaller bubble to the right on the upper side of the galaxy.
    {\bf Top right:} Same rotated H$\alpha$ narrow-band image, with MUSE pointings overlaid. Archival pointing frames are shown in orange, and GECKOS pointing frames are displayed in purple. {\bf Bottom:} A 4 colour image of MUSE data that combines H$\alpha$ (red), [\ion{O}{III}] 5007 (blue), [\ion{N}{II}]~6583 (orange), and R-band continuum (white) is shown. We use a different max and min image cut for the disk and extraplanar regions; both are log scale. There are clear filaments of ionised gas that extend over 6~kpc from the starburst of the galaxy, which connect to a central starburst region (indicated by a dashed box).}
    \label{fig:ha_bicone}
\end{figure*}

\subsection{ASKAP observations}
NGC\,4666 was observed as part of the pilot survey \citep{Westmeier2022} of Widefield ASKAP L-band Legacy All-sky Blind surveY (WALLABY). WALLABY aims at imaging the atomic hydrogen (\ion{H}{I}) in hundreds of thousands of nearby galaxies by observing the 21-cm line at 30\,arcsec resolution with the Australian Square Kilometre Array Pathfinder \citep[ASKAP,][]{Hotan2021}. The synthesised beam size is $\rm 30.1\times30.0\,arcsec^2$ ($\rm \sim2.27\times2.26\,kpc^2$) and the pixel size is $6\arcsec$ in the final data cube. As with the ALMA data, the ASKAP data was also presented in \citet{Lee2022}.

\section{Properties of the ionised gas wind of NGC\,4666 and the star formation driving it}
\label{sec:wind_def_sigma_gas}
\subsection{Extraplanar H$\alpha$ in NGC\,4666}
\label{sec:extraplanar_ha}

Figure~\ref{fig:ha_bicone} shows the deep narrow-band H$\alpha$ imaging of NGC\,4666, obtained using the Dark Energy Camera (DECam). In the left panel of the figure we label several features of the extraplanar emission that suggest feedback-driven gas expulsion. On the upper and lower side of the galaxy we identify 4 `limbs' of gas that form an X-shape and extend $\sim$5-8~kpc from the disk midplane. These are heuristically consistent with typical expectations of a biconical outflow, as emission is most clearly seen along the edges of the bicone, where the line-of-sight path length is greatest \citep[see description in][]{Veilleux2020}. 

Using the narrow-band image in Fig.~\ref{fig:ha_bicone} we measure the positions and widths of the wind components. We make these measurements using 1~arcsec wide horizontal cuts at 2.5~kpc and 5~kpc above (and below) the midplane of the galaxy. These distances are chosen because the outflow is completely distinguishable from the disk at $\pm$2.5~kpc and still has sufficient signal-to-noise for robust measurement at $\pm$5~kpc. For each horizontal cut, we identify the flux peaks on either side of the wind bicone, with the full width measured as the distance between these peaks. We find that the full width of the wind is 4.6 and 4.1~kpc on the upper and lower sides at \(\pm 2.5\)~kpc off the midplane, respectively. This increases to 6.9~kpc and 5.4~kpc at $z\sim \pm5$~kpc. The star-forming region of NGC\,4666 extends to a radius of $\sim$9-10~kpc, based on the H$\alpha$ image and Spitzer 24~$\micron$ (Fig.~\ref{fig:tdepl}). This implies that the base of the bicone covers 20-25\% of the star forming disk, but this expands to $\sim$60\% higher above the disk. This is more concentrated than \cite{McPherson2023} found in Mrk~1486, in which the base of the outflow extended far in the disk. We note, however, that Mrk~1486 is much further away (redshift $\sim0.03$) and had much lower spatial resolution.

Using the widths of the wind at 2.5~kpc and 5~kpc off the midplane, we estimate an opening angle of $\sim25\degr$ on the upper side of the galaxy and $\sim15\degr$ on the lower side of the galaxy. These are similar to the opening angle of ionised gas in M\,82 \citep{ShopbellBlandHawthorn1998}, Mrk\,1486 \citep{McPherson2023} and the wind in SBS\,0335-052E \citep{Herenz2023}. These galaxies span multiple orders of magnitude in mass (between $6\times10^6~M_{\odot}$ and $1\times10^{10}~M_{\odot}$) and exhibit a range of visual morphologies. This, however, does not seem to have large impacts on the opening angle of the outflow.  

Similar to other outflows (M\,82: \citealt{Martini2018}, NGC\,253: \citealt{Westmoquette2011}, Mrk\,1486: \citealt{McPherson2023}) the morphology of the gas in NGC\,4666 is more similar to a bi-frustrum than a pure bicone. This is because the limbs do not connect at the galaxy centre, but rather at positions offset from it, likely tracing the locations of intense star formation in the disk. This is consistent with the picture in which the wind of NGC\,4666 is driven primarily by star formation. Below we discuss the star formation properties of the disk. There is indeed a region of high SFR surface density that covers the inner $\sim$3-4~kpc of the disk, which is consistent with the region to which the limbs connect. If we assume the wind originates from the galaxy centre and force the bicone to converge there, the resulting opening angle increases to $\sim30\degr$. 

We can also measure the thickness of each of the limbs. We measure this thickness as the distance between the first flux minima on either side of each limb peak position in our horizontal flux profiles.
The thicknesses range from $\sim$1.7-2.8~kpc, and do not change significantly from the measurement near the disk or further away. 

We identify an extended region of H$\alpha$ emission on the right, upper side of the galaxy. We label this region as a ``bubble''. We note that for the following discussion it is important to recall that due to the $70\degr$ inclination of NGC\,4666 the lower side is partially blocked by the disk and oriented away from the viewer, which may impact analysis of the extraplanar morphology. In the case of a potential bubble on the lower side, this may make it more difficult to identify it.

\subsection{Star formation driving the wind}
\label{sec:sf_driving_wind}
The high inclination of NGC\,4666 introduces uncertainty in estimates of surface densities. Projection effects result in line-of-sight confusion between disk and extraplanar emission, and the deprojected physical area over which emission arises is difficult to precisely determine. As such, the surface densities presented here should be interpreted with this geometric uncertainty in mind.
\begin{figure*}
    \centering
    \includegraphics[width=1\textwidth]{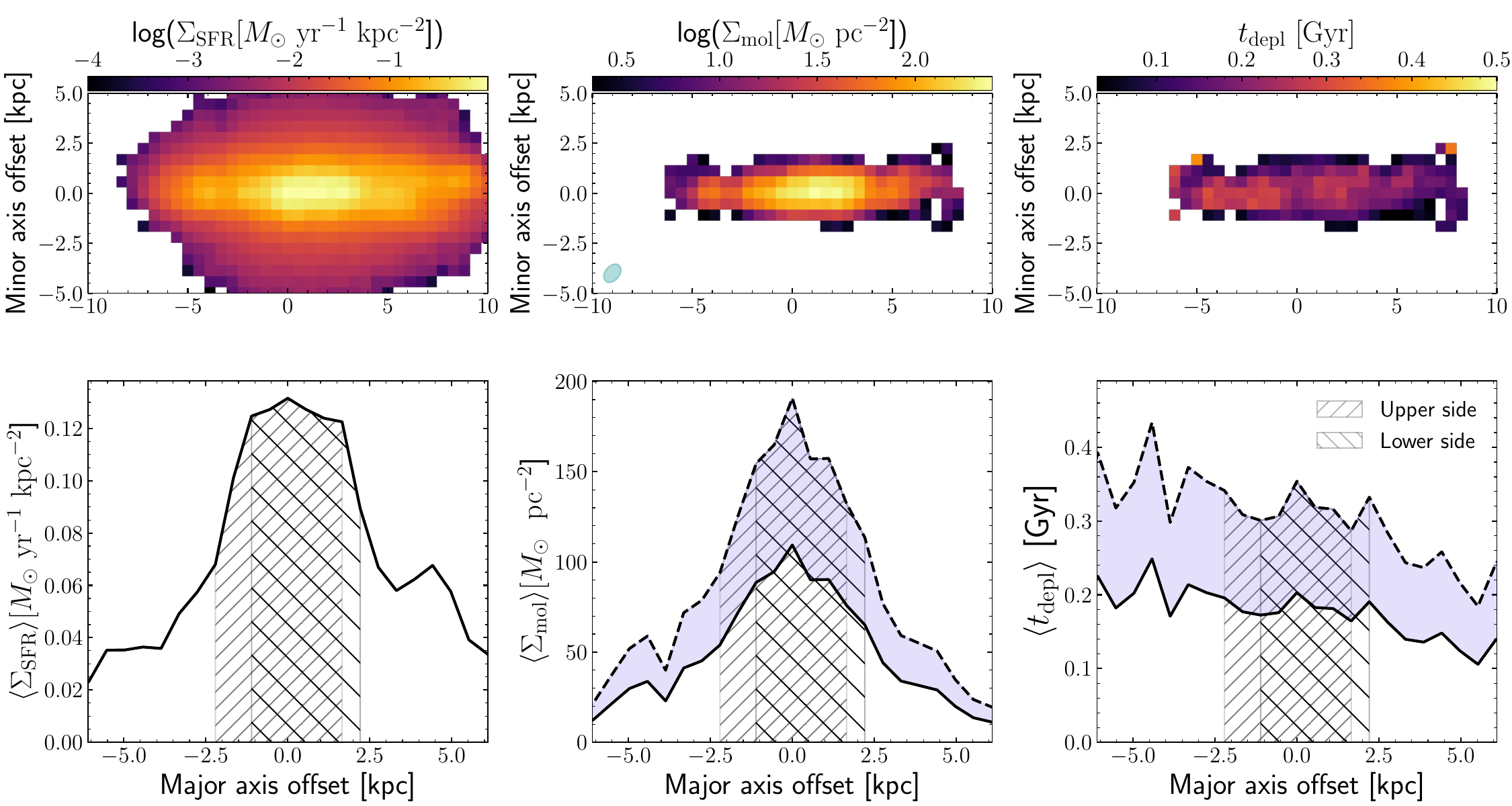}
    \caption{From left to right, the {\bf top panels} show the resolved SFR surface density, molecular gas surface density and depletion time. The ACA beam is shown in the bottom-left corner of the molecular gas surface density panel. We assume $\alpha_{\rm CO}=2.5$. The {\bf bottom panels} show the corresponding major axis radial profiles over the entire minor axis direction. For the $\Sigma_{\rm mol}$ and $t_{\rm depl}$ profiles, the solid lines correspond to a conversion factor $\alpha_{\rm CO}=2.5$, while the dashed lines refer to $\alpha_{\rm CO}=4.36$.
    The hatching represents the areas that the limbs cover at $\pm$2.5~kpc from the disk midplane (dashed box in Fig.~\ref{fig:ha_bicone}).}
    \label{fig:tdepl}
\end{figure*}
SFR surface density correlates well with the strength of star formation-driven winds \citep[reviewed in][]{Veilleux2005,Veilleux2020,ThompsonHeckman2024}. Recent resolved observations of both local Universe galaxies \citep{ReichardtChu2022b,ReichardtChu2024} and $z\sim2$ stacks \citep{Davies2019} show that this correlation is strong within galaxies. This strong connection of $\Sigma_{\rm SFR}$ with outflow strength is expected, because higher $\Sigma_{\rm SFR}$ will generate a larger number density of supernovae, which provide more energy to launch the wind. Some authors argue that a minimum threshold of SFR surface density $\Sigma_{\rm SFR}=0.1~M_{\odot}~\rm yr^{-1}~kpc^{-2}$ is required to launch a superwind \citep{Heckman2003,Newman2012}, and that SFR density determines how much material is lifted above the disk \citep{Chen2010}. The resolved observations of NGC\,4666 allow for a comparison of the properties of the disk to the location of the wind, described in the previous subsection.
To measure the star formation rate in NGC\,4666, we use the $\rm 24~\micron$ image from Spitzer/MIPS observations, binned to a spatial resolution of $\sim500$~pc. Background subtraction was performed by subtracting the median value of the data extracted in an area free of sources.
To convert the infrared flux to SFR, we use the calibration SFR$=1.31\times10^{-38}~L(\rm 24\micron)^{0.885}$, where $L(\rm 24\micron)$ is the infrared luminosity in units of erg~s$^{-1}$, and SFR is in units of $M_{\odot}~\rm yr^{-1}$ \citep{Calzetti2007}. We note that 24\micron emission can include contributions from older stellar populations, which we do not account for here. However, since NGC\,4666 is a starburst galaxy, we expect this contribution to be subdominant \citep{Temi2009,Davis2014}.

To understand the relationship between star formation and the wind, we also need to characterise the molecular gas reservoir that fuels star formation. We measure the resolved properties of the molecular gas using the ALMA/ACA moment~0 map of the data cube, regridded to match the Spitzer/MIPS data. In this subsection, we focus exclusively on the gas in the disk of NGC\,4666; we will discuss any extraplanar CO later in this work. Accordingly, the discussion of the CO-to-H$_2$ conversion factor ($\alpha_{\rm CO}$) here pertains only to the gas in the disk.
To convert CO emission to molecular gas mass, we consider two scenarios: the constant Milky Way conversion factor $\alpha_{\rm CO, MW}=4.36~M_{\odot}~(\rm K~km~s^{-1}~pc^{-2})^{-1}$ \citep{Carleton2017}; and a variable conversion factor that scales with the baryonic surface density as $\alpha_{\rm CO}\propto \Sigma_{\rm baryon}^{-0.5}$ following \citet{Bolatto2013a}
\citep[also see discussion in][]{Narayanan2011}.
Applying the latter to NGC\,4666 yields $\alpha_{\rm CO}\sim2.-2.5~M_{\odot}~\rm(K~km~s^{-1}~pc^{-2})^{-1}$ within the central $\pm1$~kpc of the disk. In the outer parts of the disk, however, a Milky Way like $\alpha_{\rm CO}$ is more appropriate. This range introduces a systematic uncertainty of $\sim$0.3~dex in our derived gas surface density and depletion time ($t_{\rm depl}$). To capture this, we adopt both $\alpha_{\rm CO}=2.5~M_{\odot}~\rm(K~km~s^{-1}~pc^{-2})^{-1}$ and $\alpha_{\rm CO, MW}$ for our analysis of the star formation in the disk.
In Figure~\ref{fig:tdepl} we show the SFR and molecular gas surface density, as well as the depletion time maps. We also display the radially averaged profiles of $\Sigma_{\rm SFR}$, $\Sigma_{\rm mol}$ and $t_{\rm depl}$. In the profiles we highlight the area that the limbs cover at $\pm$2.5~kpc from the disk midplane. We can, therefore, identify differences in the disk from the region driving the wind. 
The averaged star formation rate density within the central 5~kpc, corresponding to the region where the wind is launched, is slightly greater than $0.1~M_{\odot}~\rm yr^{-1}~kpc^{-2}$, which supports a feedback-driven superwind. The SFR surface density drops to $\sim0.06~M_{\odot}~\rm yr^{-1}~kpc^{-2}$ at a distance of $\sim3.5$~kpc from the centre. There is a small increase at $\sim4.5$~kpc that corresponds to the location of the bubble identified to the right on the upper side of the galaxy (see Figure~\ref{fig:ha_bicone}).
The highest $\Sigma_{\rm SFR}$ values of NGC\,4666 are lower than those from other well-known outflow galaxies in the local Universe (NGC\,253, M\,82 and NGC\,1482), which are all less massive than NGC\,4666.
The molecular gas surface density is slightly lower than $100~M_{\odot}\rm~pc^{-2}$ within the innermost regions of the galaxy. Outside of the central 5~kpc, it drops below $50~M_{\odot}\rm~pc^{-2}$.
Based on these two surface densities, we computed the depletion time $t_{\rm depl}$ as $t_{\rm depl}=\Sigma_{\rm mol}/\Sigma_{\rm SFR}$. This ratio corresponds to the duration required for star formation to deplete the molecular gas reservoir.
We find that in NGC\,4666 $t_{\rm depl}$ appears constant throughout the disk. The global, galaxy-averaged, depletion time is $\sim0.2~$Gyr. 
If we assume a Milky Way conversion factor, $\Sigma_{\rm mol}$ reaches $170~M_{\odot}\rm~pc^{-2}$ and goes below $100~M_{\odot}\rm~pc^{-2}$ outside of the inner 5~kpc. The galaxy-averaged depletion time in that case would be $\sim0.3~$Gyr.

\begin{figure}
    \centering
    \includegraphics[width=\columnwidth]{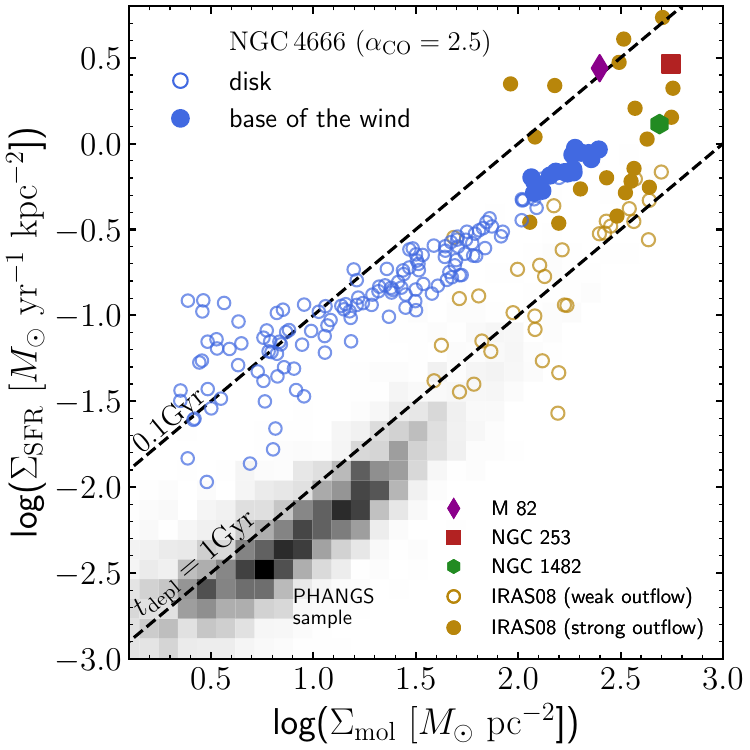}
    \caption{The resolved Kennicutt-Schmidt relationship. The measurements of NGC\,4666 are shown as blue circles: the filled points are those measured within the wind launching area (dashed box in Fig.~\ref{fig:ha_bicone}), while the empty points are the other measurements in the disk. The dark golden points correspond to similar spatial-scale measurements of IRAS\,08339+6517, presented in \citet{ReichardtChu2022b}. The grey points represent data from local face-on spirals from the PHANGS sample \citep{Sun2023}. M\,82 \citep{Leroy2015}, NGC\,253 \citep{Bolatto2013b} and NGC\,1482 \citep{VeilleuxRupke2002,Salak2020} are represented as a purple diamond, a red square and a green hexagon, respectively. The dashed lines indicate where the depletion time $t_{\rm depl}$ would be 1.0~Gyr (bottom line) and 0.1~Gyr (top line).} 
    \label{fig:ks_law}
\end{figure}

The resolved relationship between $\Sigma_{\rm SFR}$ and $\Sigma_{\rm mol}$ in galaxies, known as the Kennicutt-Schmidt law, has been extensively explored through observations \citep{Bigiel2008,KennicuttEvans2012,Leroy2013}, along with the theories that explain this connection \citep{Ostriker2010,HaywardHopkins2017,Krumholz2018}. In Figure~\ref{fig:ks_law}, we display the relation between the star formation rate density, $\Sigma_{\rm SFR}$, and molecular gas surface density, $\Sigma_{\rm mol}$, for NGC\,4666 at a sampling scale of $\rm\sim552~pc$, corresponding to the spatial resolution of the Spitzer/MIPS observations. The measurements from the region associated to the launching site of the wind show the highest $\Sigma_{\rm SFR}$ and $\Sigma_{\rm mol}$ values. They do not quite reach the values from other nearby well-known starburst-driven outflow galaxies (M\,82, NGC\,253, and NGC\,1482), but they are comparable to strong outflow measurements at similar spatial resolution in IRAS\,08339+6517, a local starburst galaxy \citep{ReichardtChu2022b}.

\section{Kinematics sub-structure of the wind}
\label{sec:wind_kinematics}
\begin{figure}
    \centering
    \includegraphics[width=1.\linewidth]{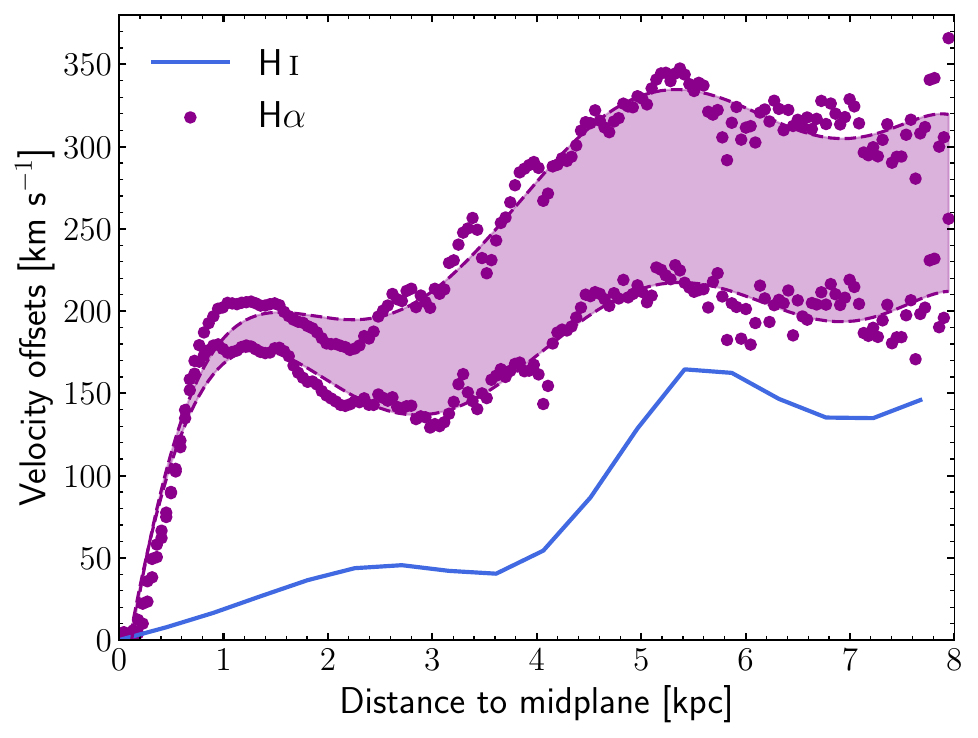}
    \caption{Gas velocity offsets within the upper cone of the NGC\,4666 superwind. The dark magenta points denote the H$\alpha$ velocity measurements from VLT/MUSE, by considering the whole outflow pointing (lower points) and the upper right main filament only (upper points). The dashed lines represent the 7th-order polynomial fits to these velocities. The blue curve represents the \ion{H}{I} velocity measurements from the WALLABY observations. The velocity values have been de-projected by dividing the observed velocities by cos($i$) (with $i=69.6\degr$ the inclination of NGC\,4666).}
    \label{fig:velocity_offsets}
\end{figure}

In Fig.~\ref{fig:velocity_offsets} we show the vertical velocity offset profiles of ionised and atomic gas. The velocity is calculated as the difference from the velocity at the midplane (defined as the average gas velocity at $z=0$ within the wind region). Note that these velocities have been corrected for the inclination of the galaxy. We adopt the photometrically-derived inclination of $69.6\degr$ rather than the kinematically-derived value of $\sim78\degr$ from \citet{Voigtlander2013}, as the former provides more conservative velocity estimates. This difference in inclination introduces a systematic uncertainty of 0.23~dex in gas velocities and all dependent quantities.
We note that our analysis assumes that the wind kinematics is dominated by vertical motions, justifying our use of cos($i$) to de-project the observed velocities. Here, we focus on the vertical component as the dominant contributor to the observed velocity offsets. Realistically, however, the observed velocities likely arise from a combination of both rotation and vertical motions.
As we show in Appendix \ref{app:gas_maps} the velocity map indicates that the H$\alpha$ filament extending upward does not rotate with the disk. This is also clear from the steep rise in H$\alpha$ velocity from the disk midplane to z$~\sim1$~kpc. After this rise $\Delta V$ remains relatively flat until z$~\sim2-3$~kpc, after which it rises again to a velocity of $\sim$250-300~km~s$^{-1}$. These velocities are comparable to typical $v_{\rm out}$ of ionised gas measurements from \citet{ReichardtChu2024}, who measure scaling relations of resolved outflow properties for 10 starburst galaxies with low inclinations from the DUVET sample. For a $\Sigma_{\rm SFR}\sim0.3$~M$_{\odot}$~yr$^{-1}$~kpc$^{-2}$, which is similar to the central $\Sigma_{\rm SFR}$ in NGC\,4666, \cite{ReichardtChu2024} predict an ionised gas outflow velocity of 240–260~km~s$^{-1}$ based on their scaling relation. These velocities are similar to the ionised gas in nearby outflows of NGC\,253 \citep{Westmoquette2011} and NGC\,1482 \citep{VeilleuxRupke2002}. The ionised gas velocity of NGC\,4666 is, therefore, comparable to other winds of similar star formation activity.

We note that the exact value of the wind velocity depends on the spatial region selected for calculation. This is shown as the shaded region in Fig.~\ref{fig:velocity_offsets}. When considering the entire GECKOS outflow pointing region, the derived velocities correspond to the lower purple points in the figure, with $v_{\rm out,H\alpha}\sim200~\rm km~s^{-1}$ at a distance of $\sim8\rm~kpc$.
Additionally, we consider just the wind filament region. This is defined as the area that exhibits continuous H$\alpha$ emission, extending up to the edge of the MUSE coverage, corresponding to a 2.7~kpc-wide section that reaches 8~kpc above the midplane. This region is indicated by a dashed grey rectangle in the left panel of Fig.~\,\ref{fig:ha_flux_vel_maps} and its different gas properties derived from the GECKOS observations are shown in Fig.~\,\ref{fig:filament_maps}.
The velocity profile for this region is represented by the upper dark magenta points in Fig.~\ref{fig:velocity_offsets}, with $v_{\rm out,H\alpha}\sim300~\rm km~s^{-1}$ at $\sim8\rm~kpc$. We decide to use the filament velocity for our analysis.

In Fig.~\ref{fig:velocity_offsets} we also show the velocity offset of the \ion{H}{I} gas. We find that the ionised gas, traced by H$\alpha$, has a velocity that is roughly twice that of \ion{H}{I}. This is similar to observations in M\,82 \citep{ShopbellBlandHawthorn1998, Martini2018}, where $v_{\rm out, H\alpha}\sim630~\rm km~s^{-1}$ and $v_{\rm out, \ion{H}{I}}\sim200~\rm km~s^{-1}$. We note that the shape and locations of rises in the \ion{H}{I} profile are similar to those in the H$\alpha$, albeit with lower amplitude. The velocity profiles of NGC\,4666 also demonstrate that the two phases we are probing share similar kinematic sub-structures. Lower outflow velocities in the neutral gas compared to the ionised components are also frequently observed in high star formation rate systems \citep{VeilleuxRupke2002, Rupke2005}.
\citet{Fluetsch2021} analysed gas kinematics maps to perform a velocity comparison of outflow phases (ionised and neutral) in a sample of 31 galaxies. They found similar structure and gradients across the field of view for the two phases.
We note that the large \ion{H}{I} beam size ($\sim30\arcsec$ or $\sim2.3~$kpc) leads to spatial averaging that may underestimate the \ion{H}{I} outflow velocities in our observations. Additionally, the brightness-weighted nature of the measurements means that emission detected at distances beyond the beam full width half maximum (FWHM) will contain contributions from both local gas and the bright central region.
The effects of the beam and the galaxy inclination are illustrated in Fig.~\ref{fig:hi_column_profiles}, where we show a surface brightness profile measured along the major axis and then reprojected on the minor axis assuming an inclination of $69.6\degr$. This reprojected profile is well matched by a Gaussian with FWHM=3.5~kpc. This suggests that the smearing effects of the beam on the \ion{H}{I} velocity offset profile are likely negligible by $z>4-5~$kpc.
\begin{figure}
    \centering
    \includegraphics[width=1\columnwidth]{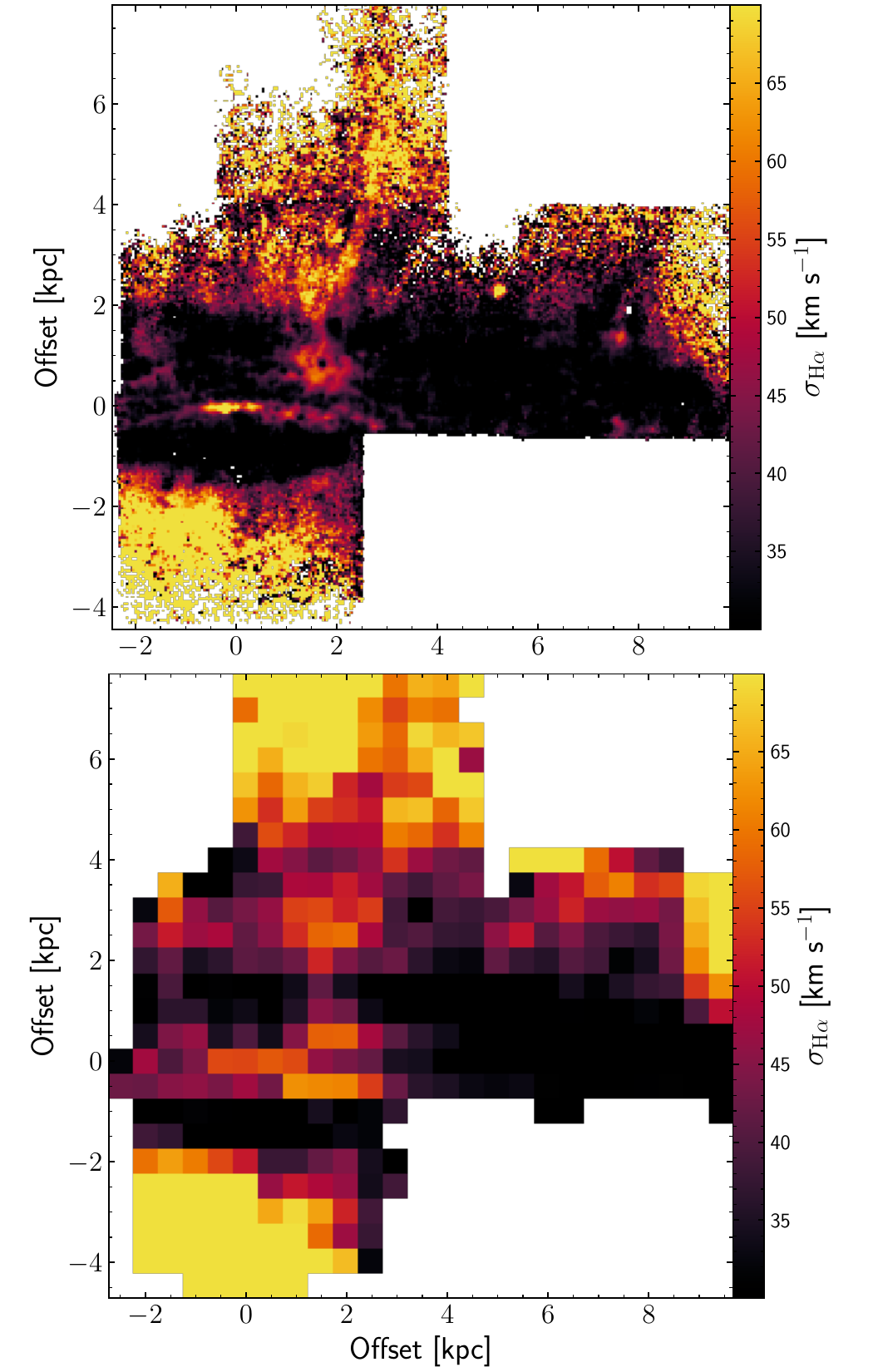}
    \caption{H$\alpha$ velocity dispersion. {\bf Top:} $\sim$45~pc spatial resolution map (0.6~$\arcsec$ pixel size). {\bf Bottom:} 500~pc spatial resolution map. The wind filament targeted by our GECKOS `outflow' pointing shows an enhanced gas velocity dispersion compared to that of the disk.}
    \label{fig:ha_sigma_and_niiha}
\end{figure}

In Fig.~\ref{fig:ha_sigma_and_niiha} we show the map of the H$\alpha$ velocity dispersion. Note that in these maps the instrumental dispersion below H$\alpha$ has been subtracted in quadrature from the velocity dispersion. A significant number of spaxels near the centre of the outflow cone are fainter, and therefore do not pass our $S/N$ criteria in the 0.6~arcsec spaxel data. We, therefore, bin the cube to a uniform sampling of 500~pc, and also remeasure the velocity dispersion in this map. This resampled cube shows a region of higher velocity dispersion toward the left side of the image. We will, therefore, use both in discussion of the velocity dispersion substructure.
We note that the observed velocity dispersion may be affected by line-of-sight projection effects in this edge-on disk, as well as by beam smearing.

Several works \citep[e.g.][]{Bik2018,McPherson2023, Watts2024} find elevated velocity dispersion gas in imaging of outflows in edge-on galaxies. The velocity dispersion in the wind of NGC\,4666 is $\sim$50-60~km~s$^{-1}$, this is similar to the velocity dispersion in the wind gas of ESO\,338-IG04 \citep{Bik2018}, but less than what is observed in Mrk\,1486 \citep{McPherson2023} and NGC\,4383 \citep{Watts2024}. We note that in both of those targets the velocity dispersion is highest near the centre of the cone. This region is not covered well in our observations. The velocity dispersion of the extended H$\alpha$ gas in the bubble, located to the right of the biconical outflow, is only mildly elevated above the disk, with an average value of $\sim$40-45~km~s$^{-1}$. However, this may be due to the lack of high altitude coverage. The velocity dispersion of gas at the same distance above the midplane over the centre of the galaxy is similar to that of the bubble. Elliott et al {\em in prep} will analyse the velocity dispersion of a subset of winds  from the GECKOS sample.

Overall, we find that the velocity dispersion of gas in NGC\,4666 is higher in the wind filament than in the main body of the disk, and is especially elevated in the wind filament above the galaxy centre. We take these observations as again similar to what is observed in galactic winds. Moreover, NGC\,4666 offers another example in which high velocity dispersion above the plane of a galaxy disk is a good indicator of outflows in galaxies.

\section{Electron density profile and Ionisation State of Wind}
\label{sec:ionised_gas_prop}
\label{sec:ne_profile}
The vertical profile of the electron density, $n_e$, is a critical prediction of galactic wind theories \citep[e.g.][]{ChevalierClegg1985,Schneider2020,FieldingBryan2022}, and is required for estimation of ionised gas mass. 

\begin{figure}
    \centering
    \includegraphics[width=1\columnwidth]{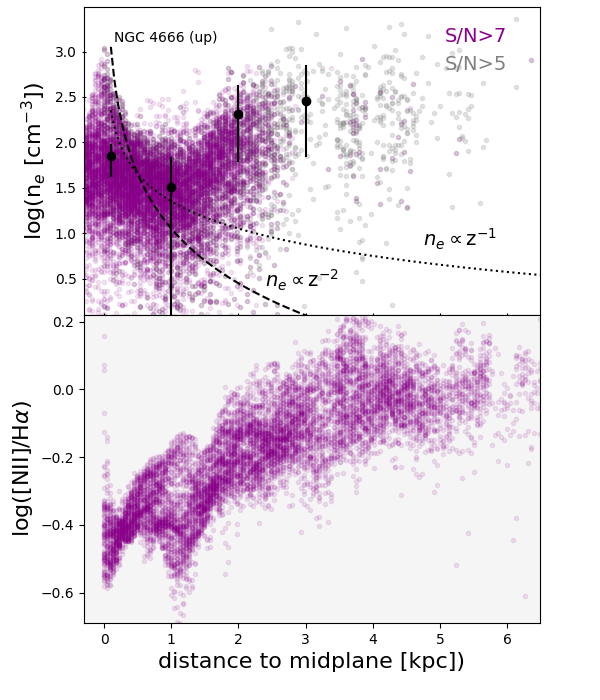}
    \caption{Profiles along the upper-right wind filament. {\bf Top:} Electron density profile derived from the [\ion{S}{II}]~$\lambda$$\lambda$6716,31 doublet. Grey symbols are spaxels with $\rm S/N>5$ and magenta symbols represent $\rm S/N>7$ spaxels. The black points show how the uncertainty on [\ion{S}{II}] flux measurements propagates to the uncertainty in electron density. The dashed and the dotted lines display electron profile decays as proposed by different outflow models. {\bf Bottom:} [\ion{N}{II}]6583/H$\alpha$ flux ratio profile.
    The ratio increases over the same distance range where the electron density rises.}
    \label{fig:ne_niiha_profiles}
\end{figure}

We estimate the electron density ($n_e$) in NGC\,4666 using the [\ion{S}{II}]~$\lambda\lambda$6716,6731 doublet. To convert this line ratio into $n_e$, we adopt the approximation from \citet{Sanders2016}, which assumes an electron temperature of $T_e\sim10^4$K. The resulting minor axis $n_e$ profile is shown in the top panel of Fig.~\ref{fig:ne_niiha_profiles}.
While GECKOS deep observations allow us to measure $n_e$, care must be taken in interpreting the results. At low electron densities ($n_e\leq 25$–30~cm$^{-3}$), this line ratio becomes essentially insensitive to $n_e$. This ``low-density limit'' varies depending on the signal-to-noise of the data \citep[see discussion in][]{Sanders2016}. Above this, the relationship between the line ratio and $n_e$ is more informative, although the slope remains shallow for densities around $n_e\sim 50$cm$^{-3}$ \citep{OsterbrockFerland2006}.

In Fig.~\ref{fig:ne_niiha_profiles} we plot all spaxels where the signal-to-noise ratio of each of the [\ion{S}{II}] lines is greater than 5 ($S/N>5$) as grey points, and highlight those with a higher ratio ($S/N>7$) in magenta. The top panel of the Fig.~\ref{fig:ne_niiha_profiles} is reproduced from Fisher et al. {\em submitted}, which studies $n_e$ for several outflow galaxies in the GECKOS sample. The black points in the figure are created to show how the flux error on the [\ion{S}{II}] emission lines propagates to the uncertainty in $n_e$.
For these black points, we compute the median flux errors on the emission lines from the spaxels in a $\Delta z\pm$0.5~kpc surrounding the position. We then propagate the uncertainty through the conversion from line ratio to electron density. Only spaxels with $S/N>7$ are included in the average. The $S/N$, however, increases systematically toward the galaxy centre.

The overall shape of the profile is not a smooth monotonic decay with distance from the galaxy. This is starkly different from expectations based on known theories of outflows \citep[e.g.][]{ChevalierClegg1985, Schneider2020, FieldingBryan2022}, which generally predict a smooth, monotonic decline in electron density with distance from the midplane. In the central 0.5~kpc of the galaxy the average is $n_e\sim100-300$~cm$^{-3}$. There is a significant number of spaxels that reach very high electron density, $n_e\sim1000$~cm$^{-3}$. We find a decline from the centre values to $\rm \sim2~kpc$. The inner declining slope of NGC\,4666 is not very dissimilar from the slope measured for M\,82 by \cite{Xu2023}, $n_e \propto z^{-1.1}$. It is important to state that the values in this region drop below the typically taken low-density limit for the electron density. We, therefore, advise a significant level of caution in interpreting the value of $n_e$ in the region from $z\sim0.5-2.5$~kpc, due to the the values of $n_e$ approaching the low-density limit. What is robust against uncertainty (since the lower uncertainty bounds do not fall below the low-density limit) is the trend: there is high $n_e$ in the galaxy centre ($>100$~cm$^{-3}$),  decline to low values between $z\sim0.5-2.5$~kpc, and a subsequent increase again beyond $\sim$2.5~kpc.

Beyond $z\sim2.5$~kpc the systematic uncertainty in $n_e$ is driven by fewer points having $S/N>7$, rather than at $z\sim0.5-2$~kpc where the high ratios of the [\ion{S}{II}] doublet generate uncertainty even at very high $S/N$.
The $n_e$ we measure, at $z\gtrsim2.5$~kpc is therefore biased to  brighter [\ion{S}{II}] emitting gas in the region.

Measurements of electron density profiles beyond $\sim$2~kpc in galactic winds remain absent from the literature. \citet{GonzalezDiaz2024} recently measure $n_e$ in extraplanar emission for $z\lesssim2$~kpc in main-sequence galaxies. While most of there $n_e$ are low, a few spaxels suggest high electron density. 
The exposure times of the GECKOS observations are designed to achieve sufficiently high signal-to-noise measurements of the \ion{[SII]}{6716,31} for measuring electron density at large distance from the disk. Fisher et al. {\em submitted} measure similar profiles of $n_e$ for six galaxies with strong outflows from the GECKOS sample, including NGC\,4666. They find that rising profiles of $n_e$ are common in the GECKOS outflows. 
\citet{Xu2023} presented an electron density profile for M\,82 to a distance of 2.5~kpc. They found an electron density of $\sim 200~\rm cm^{-3}$ at 0.5~kpc, declining to $\sim 40~\rm cm^{-3}$ at 2.5~kpc. Verna et al {\em in prep} will extend the electron density profile of M\,82 further using a wide area coverage from VIRUS-P, and likewise find high values of $n_e$. These results will help place our results for NGC\,4666 in context. \citet{Bik2018} showed a radial profile of the electron density for the starburst galaxy ESO\,338-IG04 up to a distance of 3.2~kpc. They found an electron density $n_e\sim100~\rm cm^{-3}$ in the centre and a drop to $n_e\approx10~\rm cm^{-3}$ at the largest extent of their measurements. Similarly, \cite{HamelBravo2024} report low $n_e$ in the outflow of local starburst NGC\,1569. \citet{Sharp_BlandHawthorn2010} mapped the [\ion{S}{II}]-based electron density in a sample of galaxies hosting star-formation-driven and AGN-driven winds. In the starburst NGC\,1482, they observed a density gradient, with lower values at the base of the outflow than in the outskirts. We note, however, that the measurements of \citet{Sharp_BlandHawthorn2010} do not extend far into the wind, and it is difficult to assess the shape of the profile. The GECKOS results shown here and in Fisher et al. {\em submitted} are unique in probing $n_e$ far from the galaxy.

While the shape of the electron density profile of NGC\,4666 is different from expectations in theory, it is not inconsistent with unresolved observations of electron density in outflows of ULIRGs and high-z galaxies. These so-called down-the-barrel observations collapse the entire outflow into a single broad component of the spectral line in a less inclined galaxy. \citet{Fluetsch2021} studied 26 (U)LIRGs using MUSE and found that, on average in their sample, the electron density of the ionised gas in the outflow —derived from the [\ion{S}{II}] doublet— was roughly three times higher than in the disk ($\langle n_{e,\rm outflow}\rangle\sim 490~\rm cm^{-3}$, $\langle n_{e,\rm disk}\rangle\sim 145~\rm cm^{-3}$). At higher redshift ($0.6<z<2.7$), \citet{ForsterSchreiber2019} measured a typical electron density $n_e\sim 380~\rm cm^{-3}$ in star-formation driven winds and $n_e\sim 1000~\rm cm^{-3}$ in AGN-driven winds.
Our observations of NGC\,4666 wind filament exhibit lower values than galaxies at higher redshift. They are also lower than the electron density derived in the (U)LIRGs sample.

The bottom panel of Fig.~\ref{fig:ne_niiha_profiles} presents the vertical profile of [\ion{N}{II}]6583/H$\alpha$ flux ratio derived from the MUSE data. This ratio is sensitive to the ionisation state and hardness of the radiation field, and is commonly used to trace changes in excitation conditions. All weighted average values are below log([\ion{N}{II}]/H$\alpha$)=-0.2 in the inner 2~kpc, then the ratio increases until it becomes more constant from $\sim$3~kpc onwards (log([\ion{N}{II}]/H$\alpha$)$\sim$0). These values of [\ion{N}{II}]/H$\alpha$ are not indicative of strong shocks. An interesting result is that the increase in the [\ion{N}{II}]/H$\alpha$ ratio values occurs within the same distance range as the increase in the electron density. This correlation strengthens our interpretation that these observations are driven by underlying physical processes occurring within this region (z $\sim$ 2-3~kpc).
Beyond this distance, the [\ion{N}{II}]6583/H$\alpha$ ratio remains elevated ($\sim1$) and shows little variation with height (Fig.~\ref{fig:filament_maps}), suggesting that the ionisation conditions in the extraplanar gas remain relatively uniform.
Similar enhanced line ratios in diffuse extraplanar gas have been reported by \citet{GonzalezDiaz2024} along the minor axis of their studied galaxies, and Elliott et al. {\em in prep} show that strong optical emission line ratios increase in the winds of nine star-forming GECKOS galaxies.

We summarise our results on the electron density profile in the outflow of NGC\,4666. We find that the electron density of gas decreases from the galaxy centre to the edge of the disk at $z\sim1-2$~kpc. We then find that there is high density gas in the wind. This gas has $n_e\sim100-300$~cm$^{-3}$ and resides at $z>2$~kpc. We cannot say if this represents all of the gas in the wind, or if it is only a subset of the brightest gas clouds. Moreover, it is important to state that the size of gas clouds in the wind is likely smaller than the GECKOS resolution ($\rm \sim45~pc$). Gas clouds in M\,82 wind have been found to have typical widths of $\rm \sim5-18~pc$ \citep{Fisher2025}. The electron density of those clouds may be higher than the surrounding medium, which would indicate even higher electron densities than what we measure in small regions. High-resolution observations of outflows with \textit{JWST} are a clear step towards better understanding this critical parameter. 

\section{Outflow rates in NGC\,4666}
\label{sec:outflow_rates}
In this section, we estimate mass outflow rates for ionised and \ion{H}{I} gas phases in the galactic wind. We use the DECam H$\alpha$ narrow-band observations presented in Fig~\ref{fig:ha_bicone} to define the superwind bicone.
We measure outflow mass rates of the warm ionised gas and of the neutral atomic gas using H$\alpha$ and WALLABY observations, respectively. We also derive a mass outflow rate upper limit for the molecular gas from the ALMA/ACA observations; details and calculations are presented in Sect.~\ref{sec:co_mass}.

The resolved outflow rate measurement in a cell within the outflow is defined as:
\begin{equation}
    \dot{M}_{\rm out, z} = M_{\rm out, z}~\frac{v_{\rm out, z}}{\Delta z}
\end{equation}
where $M_{\rm out, z}$ is the gas mass in the cell, $v_{\rm out, z}$ is the velocity of the outflowing gas in that cell and $\Delta z$ is the physical size of the cell.
To compare \ion{H}{I} and ionised gas outflow properties, we first convolve the H$\alpha$ narrow-band image to match the spatial resolution of the WALLABY \ion{H}{I} map. We find that the ionised gas mass outflow rate measurements show good agreement between the full resolution H$\alpha$ map and convolved data. 
For $v_{\rm out}$, we adopt the velocity offset of the wind filament presented in Fig.~\ref{fig:velocity_offsets}, derived from the GECKOS observations. We fit a 7$^{\rm th}$ order polynomial to the H$\alpha$ profile and evaluate this polynomial at the \ion{H}{I} resolution.

We define the wind region based on its position and width measurements discussed in Sect.~\ref{sec:extraplanar_ha}. It is $\sim6.9~$kpc wide at 5~kpc above the disk, being robustly detected from \(-4.8\) kpc on the left side to 2.1~kpc on the right side. The limbs are completely distinguishable up to $\sim8$kpc from the midplane. The outflow region we consider for our measurements is then a rectangular region covering $\rm \sim6.9\times8~kpc^2$.

In this wind, we calculate the resolved outflow rate measurements by summing all pixels in the horizontal direction that are within the $\Delta z$ region ($\Delta z$ corresponding to the size of a WALLABY pixel, $\sim450$~pc).

Estimation of the mass-outflow rate of ionised gas and atomic gas are both subject to different sources of systematic uncertainty. We will describe each of these estimates, and their caveats in the subsections below.

\subsubsection{Resolved ionised gas mass outflow rate}
\begin{figure}
    \centering
    \includegraphics[width=1\columnwidth]{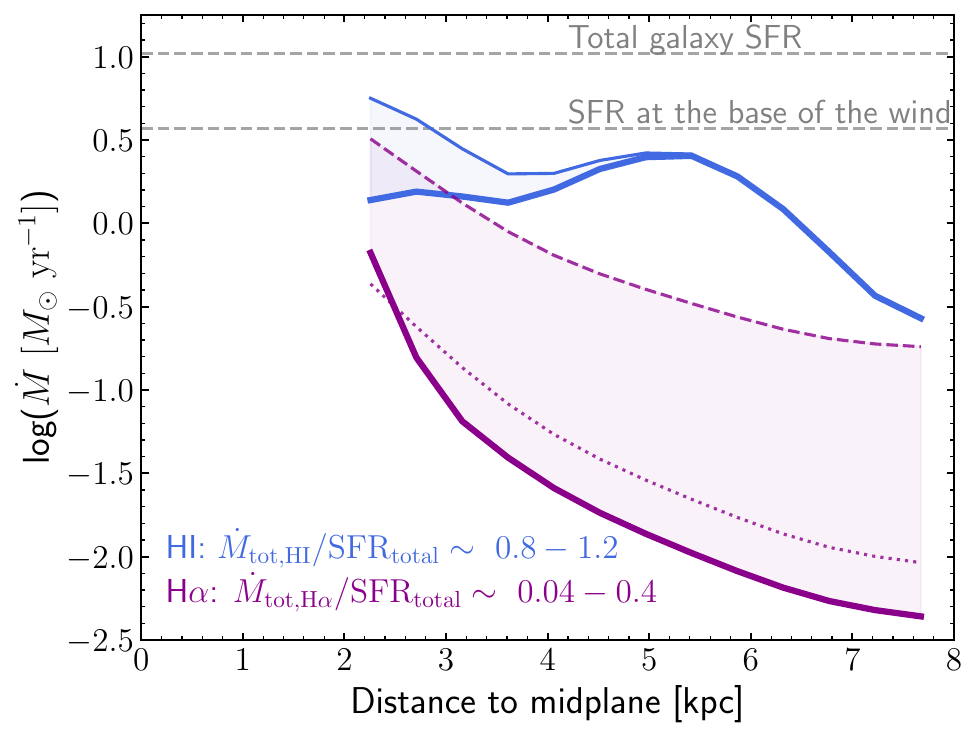}
    \caption{Mass outflow rates against distance to the midplane. The blue curves show the \ion{H}{I} mass outflow rate measurements: the upper thin line corresponds to measurements before accounting for projection effects, while the thicker line presents corrected estimates (see Sect.\,\ref{sec:hi_mdotout}).
    The magenta shaded area displays the H$\alpha$ mass outflow rate measurements, assuming a clumping factor $C_e=1$. 
    The upper magenta dashed line corresponds to the low-density assumption where $n_e$ decays as $z^{-0.8}$; the dotted line represents measurements for a constant $n_e=100~\rm cm^{-3}$; and the bottom solid line is the profile for the data-driven profile of $n_e$, given by Eq.\,\ref{eq:datadriven_ne}.
    The grey dashed lines indicate star formation rates: the bottom line shows the infrared-based SFR integrated within the region underneath the outflow, while the upper line represents the total SFR from \citet{Vargas2019}.
    Values in the bottom-left corner give total mass loading factors for each phase, calculated by summing the measurements beyond 2~kpc.}
    \label{fig:mdotout}
\end{figure}

The ionised gas mass is estimated using the extinction-corrected luminosity of H$\alpha$, $L_{\rm H\alpha}$, with the following \citep{Veilleux2020}, 
\begin{equation}
    M_{\rm ion}[M_{\odot}] = 3.3 \times 10^8 C_e \frac{L_{44}(\rm H\alpha)}{n_{e,3}}
\end{equation}
Here $n_{e,3}$ is $n_e/1000$ and $L_{44}(\rm H\alpha) = L(\rm H\alpha)/10^{44}$. $C_e$ is the electron density clumping factor, $C_e \equiv \langle n_e^2 \rangle/n_e^2$. If each cloud has uniform density then $C_e$ is of order unity.
If the gas clouds are clumpier, with more significant density variations, the clumping factor would have a higher value. Some simulations suggest that clumping factors greater than unity are realistic \citep[e.g.][]{Schneider_Robertson_2018}. In the absence of direct observational constraints, assuming $C_e=1$ is common. For the profiles shown in Fig.~\ref{fig:mdotout}, we assume $C_e=1$.
We mask measurements taken at distances smaller than the \ion{H}{I} beam major axis length ($\sim 2.26~$kpc).
We also must adopt a prescription for the electron density, $n_e$. Based on the profiles displayed in Fig.~\ref{fig:ne_niiha_profiles}, we choose to make multiple assumptions, which allows us to track this systematic uncertainty through our main results. The first is a ``data-driven'' assumption, which assumes a profile that is determined from the data described in Sect.~\ref{sec:ne_profile}. We derive the following equation using the \texttt{python} package ODR\footnote{https://docs.scipy.org/doc/scipy/reference/odr.html}:
\begin{equation}
    n_e\ [{\rm cm^{-3}}] \sim \begin{cases}
        45~\left(\frac{z}{0.5~\rm kpc}\right)^{-0.8} & z \leq 2~{\rm kpc} \\
        195 {z} - 375 & 2~{\rm kpc} < z \leq 3~{\rm kpc} \\
        210 & z > 3~{\rm kpc}
\end{cases}
\label{eq:datadriven_ne}
\end{equation}

The second assumption for $n_e$ is a ``decaying'' assumption, where $n_e$ decays $\propto z^{-0.8}$ across the entire distance range. This assumption leads to values of order $\rm \sim10~cm^{-3}$ for the highest altitudes ($z~\sim3-7~$kpc), and is similar to what has been assumed historically \citep{ShopbellBlandHawthorn1998}. Third, we consider a constant value of $n_e=100~\rm cm^{-3}$, which is similar to other works in the literature \citep[e.g.][]{ReichardtChu2024}.

\label{sec:ion_mdotout}
Based on the electron density profile derived directly from the data (top panel of Fig.\,\ref{fig:ne_niiha_profiles}, Eq.\,\ref{eq:datadriven_ne}), we estimate a total ionised mass outflow rate of $\sim0.5~M_{\odot}~\rm yr^{-1}$. This profile is considered our preferred solution, as it is directly constrained by observations.
For comparison, adopting a steeper electron density profile that decays as
$\propto z^{-0.8}$ up to 8~kpc yields a higher mass outflow rate of $\sim4.7~M_{\odot}~\rm yr^{-1}$.
We estimate an ionised mass outflow rate of $\sim0.5~M_{\odot}~\rm yr^{-1}$ for the data-driven approach. The assumption of constant $n_e=100~\rm cm^{-3}$ yields a result of $\sim0.5~M_{\odot}~\rm yr^{-1}$. The largest difference is the decaying profile, which is nearly an order of magnitude higher at $\sim4.7~M_{\odot}~\rm yr^{-1}$.
We also note that the decay in mass varies based on the assumption of $n_e(z)$. We find for the data-driven that $M_{\rm out, ions}$ declines by a factor of 50 between 0.5~kpc and 3~kpc. Conversely, the assumption of a decaying profile leads to a 3x decline across a similar distance.

In M\,82, mass outflow rate estimates derived from H$\alpha$ observations span values from $\sim6.3~M_{\odot}~\rm yr^{-1}$ at 0.7~kpc to $\sim1~M_{\odot}~\rm yr^{-1}$ at 2.2~kpc \citep{Xu2023, ShopbellBlandHawthorn1998}.
\cite{Xu2023} used average electron density values in radial bins that decline from $\sim$200~cm$^{-3}$ at 0.5~kpc to $\sim$40~cm$^{-3}$ at 2.2~kpc, approximating the power law $n_e\sim100\left(\frac{z}{1165~\rm pc}\right)^{-1.17}$. If we compare this to the decaying electron density profile in our results, then the ionised gas mass loading is similar.
Our data-driven electron density profile, however, at larger distances from the disk.
Verna et al {\em in prep} find similar results of high $n_e$ at large radius in M\,82 using VIRUS-P observations, this has a similar impact on the mass profile as in NGC\,4666.
In NGC\,1482, the ionised gas mass outflow rate is $\sim0.6~M_{\odot}$~yr$^{-1}$. This value is calculated for an electron density $n_e\sim$10~cm$^{-3}$ \citep{Sharp_BlandHawthorn2010}, which is measured on the [\ion{S}{II}] doublet in observations that do not reach the distances of our measurement in NGC\,4666. We, therefore, cannot say if NGC\,1482 has a rise in $n_e$ at $z~>2$~kpc similar to NGC\,4666.

\begin{figure}
    \centering
    \includegraphics[width=\columnwidth]{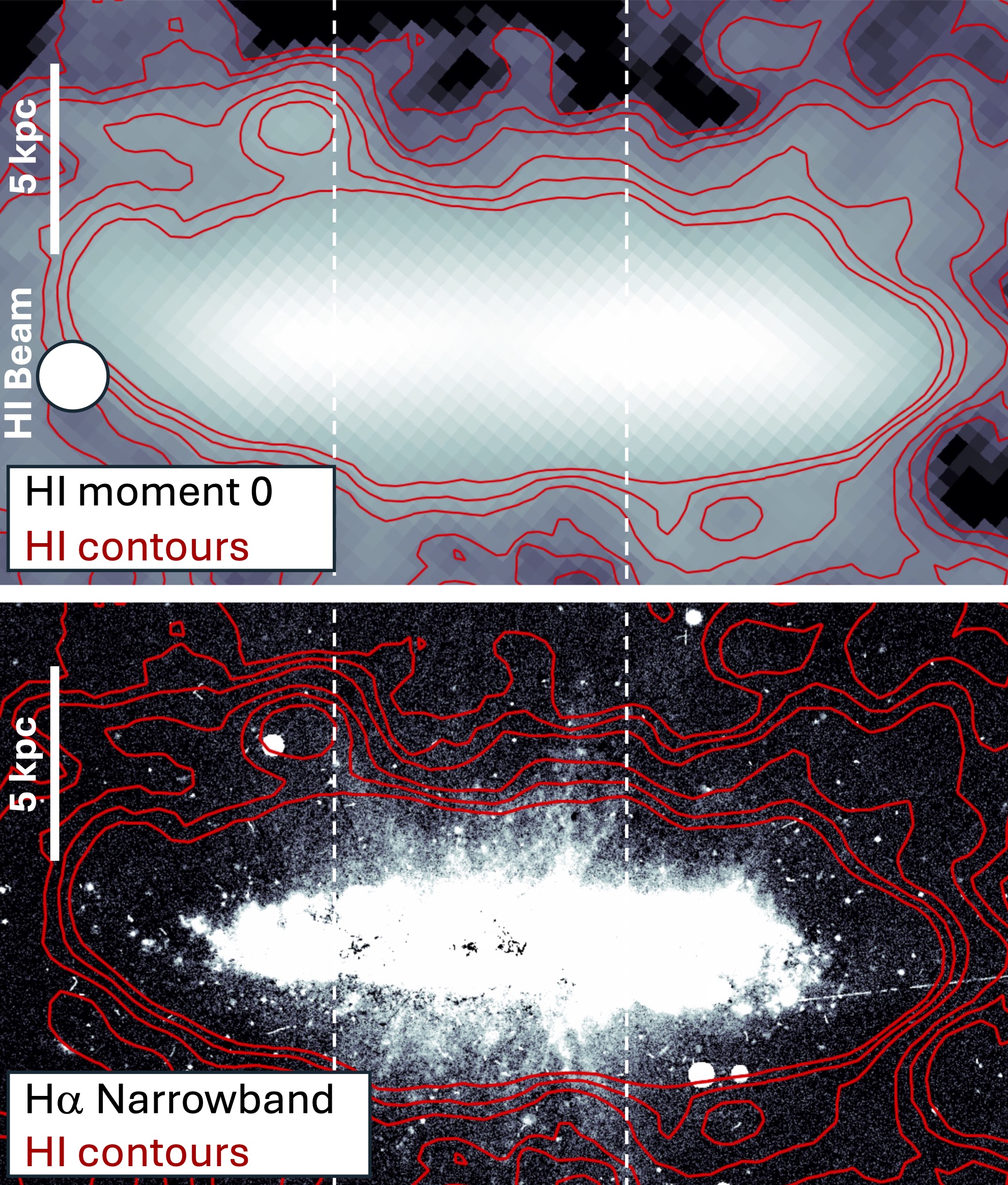}
    \caption{\textbf{Top panel} shows the \ion{H}{I} moment 0 map from \citet{Lee2022}, with red contours for \ion{H}{I} gas. The contour levels are chosen to guide the eye to the similarity of H$\alpha$ and \ion{H}{I}. \textbf{The bottom panel} shows the H$\alpha$ narrowband image with \ion{H}{I} contours overlaid. In both panels there is a dashed vertical line that is set to guide the eye. The bottom panel shows that morphology of the \ion{H}{I} contours correlates with the location of the H$\alpha$, which can be interpreted as the \ion{H}{I} gas being impacted by the wind.}
    \label{fig:HaHI}
\end{figure}   

\begin{figure}
    \centering
    \includegraphics[width=\columnwidth]{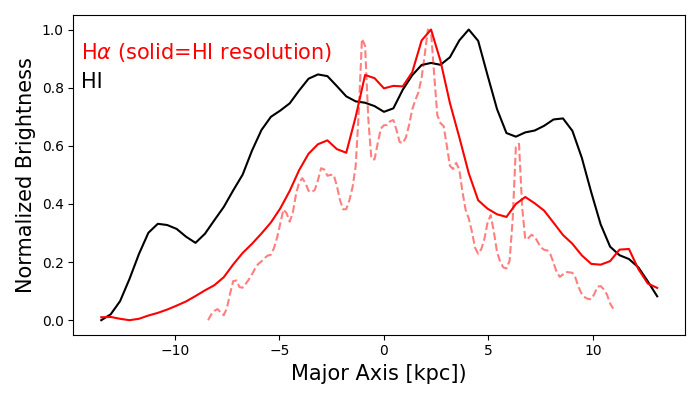}
    \caption{Horizontal cuts to the H$\alpha$ (red) and \ion{H}{I} (black) images. The cuts are made at 2.5~kpc above the disk, and extend upward to 5~kpc. Profiles are normalized such that the maximum surface brightness along the profile is one and the minimum is zero. For H$\alpha$ we show both the full resolution and the image that has been blurred to match the \ion{H}{I} resolution.}
    \label{fig:cuts}
\end{figure}

\subsubsection{Resolved \ion{H}{I} gas mass outflow rate}
\label{sec:hi_mdotout}
The \ion{H}{I} gas mass is calculated from the velocity-integrated intensity (moment 0 map). While the conversion of mass in \ion{H}{I} is well defined, the largest systematic arises from our ability to separate the \ion{H}{I} gas in the outflow from that of surrounding gas both in the disk and extraplanar emission. We have shown above that the ionised gas in NGC\,4666 has observed morphology and kinematics that are consistent with galactic winds. Moreover, observations of very nearby galaxies like M\,82 \citep{Martini2018} and NGC\,1569 \citep{Johnson2012} suggest that ionised gas outflows are likely to contain \ion{H}{I} outflow as well. We, therefore, consider the scenario in which some fraction of the extraplanar \ion{H}{I} near the H$\alpha$ outflow is also outflow gas. Nonetheless, \citet{Lee2022} shows that a significant amount of gas around NGC\,4666 is due to interactions within the group. In this subsection we will attempt to place constraints on the mass of the \ion{H}{I} wind, with these uncertainties in mind. 

In Fig.~\ref{fig:HaHI} we compare the morphology of the extraplanar \ion{H}{I} gas to the H$\alpha$ gas. In the H$\alpha$ map biconical filaments of ionised gas extend upward from the disk with a minimum in emission that is above the centre of the galaxy, as described above in this work. This is consistent with the expectations of a multiphase wind in which the hot inner cone is surrounded by 10$^4$~K ionised gas. \citep{Heckman2003,Veilleux2005,Veilleux2020,Bolatto2013b,Leroy2015}. The contours of \ion{H}{I} gas show an outward bulge that is correlated to the location of the H$\alpha$ gas outflow limbs, and likewise a minimum in that bulge that is colocated with the minimum in H$\alpha$. The dashed lines in the figure are set to guide the eye and show this bulge. In Fig.~\ref{fig:cuts} we show horizontal surface brightness profiles of \ion{H}{I} and H$\alpha$ for gas at 2.5-5~kpc above the midplane, on the upper side of the galaxy. These further illustrate the connection between the bicones in H$\alpha$ with increases in \ion{H}{I} surface brightness. The \ion{H}{I} peaks located near to -3~kpc and 4~kpc correspond to the bulges in the contours above the disk plane in Fig.~\ref{fig:HaHI}. The ionised gas is peaked inside of these \ion{H}{I} peaks. This nested structure in which the \ion{H}{I} peaks surround the ionised gas is consistent with the standard expectation of multiphase outflows, illustrated well in Fig.~19 of \citet{Leroy2015}. While we cannot know for certain what causes multiphase morphological changes, it is, nonetheless, straightforward to interpret these morphological correlations as having a causal connection. In this case one interpretation of the changes in the \ion{H}{I} morphology over the centre of the galaxy is that the gas is connected to the outflow observed in H$\alpha$.

A significant source of uncertainty is that extraplanar \ion{H}{I} emission may be due to both gas in the outflow and gas that is due to the interaction \citep[described in][]{Lee2022}. Our assumption that the \ion{H}{I} gas that is colocated in the image with H$\alpha$ bicone is likely an oversimplification. It is difficult to estimate the difference, as any interaction will be very asymmetric. Readers should take these numbers as an upper estimate on the outflow mass-loading in cold atomic gas. 

In NGC\,4666 we estimate a total atomic mass of the outflow of $M_{\rm out}=4\times10^8~M_{\odot}$, measured from 2~kpc above the midplane, which is 8\% of the total \ion{H}{I} mass in the galaxy. We estimate a mass outflow rate of $13~M_{\odot}~\rm yr^{-1}$. We find that the mass outflow rate declines by roughly an order of magnitude from the region near the galaxy compared to those at $z=6-7~\rm kpc$. This is due to the decrease in the mass surface density at higher altitude. We find that for \ion{H}{I} gas above 2~kpc, 90\% is contained within the region from 2-5~kpc.

Another source of systematic uncertainty on the mass of \ion{H}{I} in the wind is from the large beam of the WALLABY data (FWHM$\sim2.2$~kpc). Some gas could be projected into the wind from the disk. To estimate this uncertainty we model the minor-axis profile by assuming a tilted disk that is completely symmetric. This is likely a simple description of the galaxy, but sufficient for our purposes.
We measure the surface brightness profile along the major-axis and then re-project that profile on the minor-axis, assuming an inclination of $69.6\degr$. We find that the minor axis disk profile can be described by a Gaussian with FWHM$\sim3.5$~kpc. We then project such a profile into the outflow and subtract the emission from the observed \ion{H}{I} brightness in each pixel.
This is described in  Appendix~\ref{app:inclination_effect}, illustrated in Fig~\ref{fig:hi_column_profiles}.
This correction affects the \ion{H}{I} mass estimates in the wind: when integrating beyond 2~kpc, it introduces a total uncertainty on the outflow mass of $\sim$0.3~dex.
The impact is more significant between 2 and 4~kpc to the midplane, where applying this correction decreases the \ion{H}{I} mass by $\sim61\%$.

It is not as straightforward to estimate the impact of extraplanar emission from the group interactions. We could assume that group interactions become more important further from the galaxy. If we further remove all gas beyond 5~kpc from the wind with our disk-blurring correction, this would decrease the mass outflow rate by an additional 37\%.

Overall the mass outflow rate of \ion{H}{I} is not likely larger than 13~$M_{\odot}~\rm yr^{-1}$. If we, however, correct both for the blurring of the disk and the gas from the group the mass outflow rate could be as low as $5.4~M_{\odot}~\rm yr^{-1}$.

For comparison, in M\,82 the mass outflow rate of \ion{H}{I} is of order $\sim$3~$M_{\odot}~\rm yr^{-1}$. There is an order of magnitude decrease in the mass outflow rate from a distance of $\sim$1~kpc to $\sim$5~kpc. In both of these targets the lower velocity of \ion{H}{I} implies that the dominant mass component may not be travelling as far from the disk. 

There are very few examples of outflow rates derived from \ion{H}{I} 21~cm emission in the literature. \citet{RobertsBorsani2020} attempted to detect broad emission in \ion{H}{I} 21~cm stacked spectra from multiple surveys but did not find anything significant. Their upper limit on the atomic mass outflow rate is $\dot{M}_{\rm \ion{H}{I}}<26.72~M_{\odot}~\rm yr^{-1}$.
Comparisons can be made with indirect neutral gas tracers such as \ion{Na}{D} and [\ion{C}{II}]. However, it is important to note that these tracers do not probe the same gas phase: \ion{Na}{D} traces warmer gas than \ion{H}{I}, and [\ion{C}{II}] is subject to systematic uncertainties that make it difficult to interpret for outflows. For context, \citet{RobertsBorsani2020} also re-derived neutral outflow rates from the fitted parameters of \ion{Na}{D} profiles given by \citet{Krug2010} for a sample of 35 infrared-faint Seyfert galaxies. They found maximum values of $\sim 23-56~M_{\odot}~\rm yr^{-1}$ for galaxies above the main sequence.

\subsubsection{Multiphase mass loading of NGC\,4666}
\label{sec:mass_loading}
The lower grey dashed line in Fig.~\ref{fig:mdotout} represents the infrared-based star formation rate of the region underneath the wind. We compute this SFR$_{\rm central}$ from the Spitzer/MIPS $24\micron$ observations and find ${\rm SFR_{central}}=3.7~M_{\odot}~\rm yr^{-1}$. The upper grey dashed line shows the total galaxy SFR$_{\rm total}=10.5~M_{\odot}~\rm yr^{-1}$ \citep{Vargas2019}.
The mass loading factor, $\eta_{\rm M}=\dot{M}_{\rm out}/\rm SFR$, describes the relationship between the rate of gas expelled from a region of the galaxy and the star formation activity responsible for driving the outflow. We calculate a total mass loading factor $\eta_{\rm M}$ in each phase where we sum the mass outflow rate measurements beyond 2~kpc, using both the central SFR and the total galaxy SFR. As we have argued previously in this work, the main biconical wind in NGC\,4666 is likely driven by a centrally located starburst, however, unresolved observations of outflows do not make this distinction. We will therefore provide both the mass loading of the central SFR and the global, so that our results can be directly compared to previous works.
We assume that NGC\,4666 wind is symmetric and multiply the mass outflow rates derived from our observations by two to account for the lower side of the wind.

For the atomic gas, we measure a total mass outflow rate $\dot{M}_{\rm out,\ion{H}{I}}\sim13~M_{\odot}~\rm yr^{-1}$ (see Sect.\,\ref{sec:hi_mdotout}). This corresponds to mass loading factors of $\eta_{\rm M, \ion{H}{I}}\sim3.5$ (using SFR$_{\rm central}$) or $\eta_{\rm M, \ion{H}{I}}\sim1.2$ (using SFR$_{\rm total}$). As discussed in the previous subsection, beam correction effects reduce the \ion{H}{I} mass outflow rate to 8.5$~M_{\odot}~\rm yr^{-1}$, yielding corrected mass loading factors of $\eta_{\rm \ion{H}{I}}\sim2.3$ (with SFR$_{\rm central}$) or $\eta_{\rm M, \ion{H}{I}}\sim0.8$ (with SFR$_{\rm total}$).

For the warm ionised phase, we measure $\dot{M}_{\rm out,H\alpha}$ between 0.5 and 4.7$~M_{\odot}~\rm yr^{-1}$ (see Sect.\,\ref{sec:ion_mdotout}), depending on our electron density assumptions. This corresponds to mass loading factors $\eta_{\rm M, H\alpha}$ between 0.1-1.3 (using SFR$_{\rm central}$) or 0.04-0.4 (using SFR$_{\rm total}$). Using the electron density profile measured from the GECKOS data gives values at the lower end of these ranges.

Under all estimates of the \ion{H}{I} mass loading, even the most conservative, the atomic phase dominates the mass outflow rate of the ionised gas by a significant amount, roughly an order of magnitude.
This result is critically dependent on our measurement of high electron densities in the wind. Some authors have historically made assumptions of $n_e\sim10~\rm cm^{-3}$, which would have made the mass loading of the ionised gas comparable to \ion{H}{I}. Similarly, a decaying $n_e$ profile would also have resulted in a larger ionised mass outflow rate.

Before comparing to other published works, it is important to discuss systematic differences in measurement techniques. The vast majority of mass loading factors are determined with so-called ``down-the-barrel'' measurements in which the outflow properties are derived from decomposition of the emission lines in a mostly face-on target. A single velocity, which is typically similar to the 90\% velocity of the broad component, is taken to describe all of the gas.
Our measurements are likely biased toward lower mass outflow rates for multiple reasons. First, we cannot distinguish outflow gas from disk gas for $z~\lesssim1.5-2$~kpc. The velocity profile of the ionised gas presented in Fig.~\ref{fig:velocity_offsets}, however, shows that the velocity increases rapidly to $\rm \sim200~km~s^{-1}$, likely indicative that it is already tracing outflow gas. Due to the edge-on orientation, our measurements may be missing some fast-moving gas that would be included in down-the-barrel measurements of face-on galaxies. Moreover, the velocity profile rises again from 200~km~s$^{-1}$ further to 300~km~s$^{-1}$. If we take the total luminosity of H$\alpha$ in the wind of NGC\,4666, adopt $v_{\rm out}\sim250$~km~s$^{-1}$ and r$_{\rm out}\sim1$~kpc we derive an $\dot{M}_{\rm out}$ that is $\sim2-3\times$ our resolved measurement. We therefore expect that observational biases may lead to mass loading factors that are $\sim$0.3~dex lower than for face-on galaxies.

Our total ionised gas mass loading factor lies within the range, albeit on the lower side, of those from the DUVET sample \citep{ReichardtChu2024}. They presented sub-kpc, spatially-resolved measurements of mass outflow rates and found $\eta_{\rm M, ion}\sim0.1-10$. It is also consistent with the measurement of \citet{McPherson2023}, who studied the outflow in the edge-on galaxy Mrk\,1486 using optical emission line observations. They derived an ionised mass loading factor $\eta_{\rm M, ion}=0.7$ within the outflow along the minor axis. NGC\,4666 has a lower star formation rate compared to the galaxies from the DUVET sample analysed by \citet{ReichardtChu2024} and \citet{McPherson2023}. The galaxies in the DUVET sample were chosen for having star formation rates that exceed the main sequence value by at least a factor of five based on their total stellar mass. Given this, the lower ionised gas mass loading factor in our galaxy is therefore not unexpected. 

Using FIRE-2 simulations, \citet{Pandya2021} determined that warm gas ($10^3~{\rm K} < T < 10^5~{\rm K}$) mass loading factors are smaller than unity for galaxies with a stellar mass similar to NGC\,4666. This prediction is consistent with our observed ionised gas mass loading factors for NGC\,4666. Similarly, the SMAUG-TIGRESS simulations \citep{Kim2020} found mass loading factors for cool gas ($T\sim10^4~$K) ranging from $\eta_{\rm M}\sim0.2−10$ for $\Sigma_{\rm SFR}\sim0.1−1~M_{\odot}~\rm yr^{−1}~kpc^{−2}$. These surface densities are comparable to the $\Sigma_{\rm SFR}$ we derive in the disk of NGC\,4666, and their predicted mass loading factors align with our total $\eta_{\rm M, H\alpha}$ measurements.

\begin{figure}
    \centering
    \includegraphics[width=1\columnwidth]{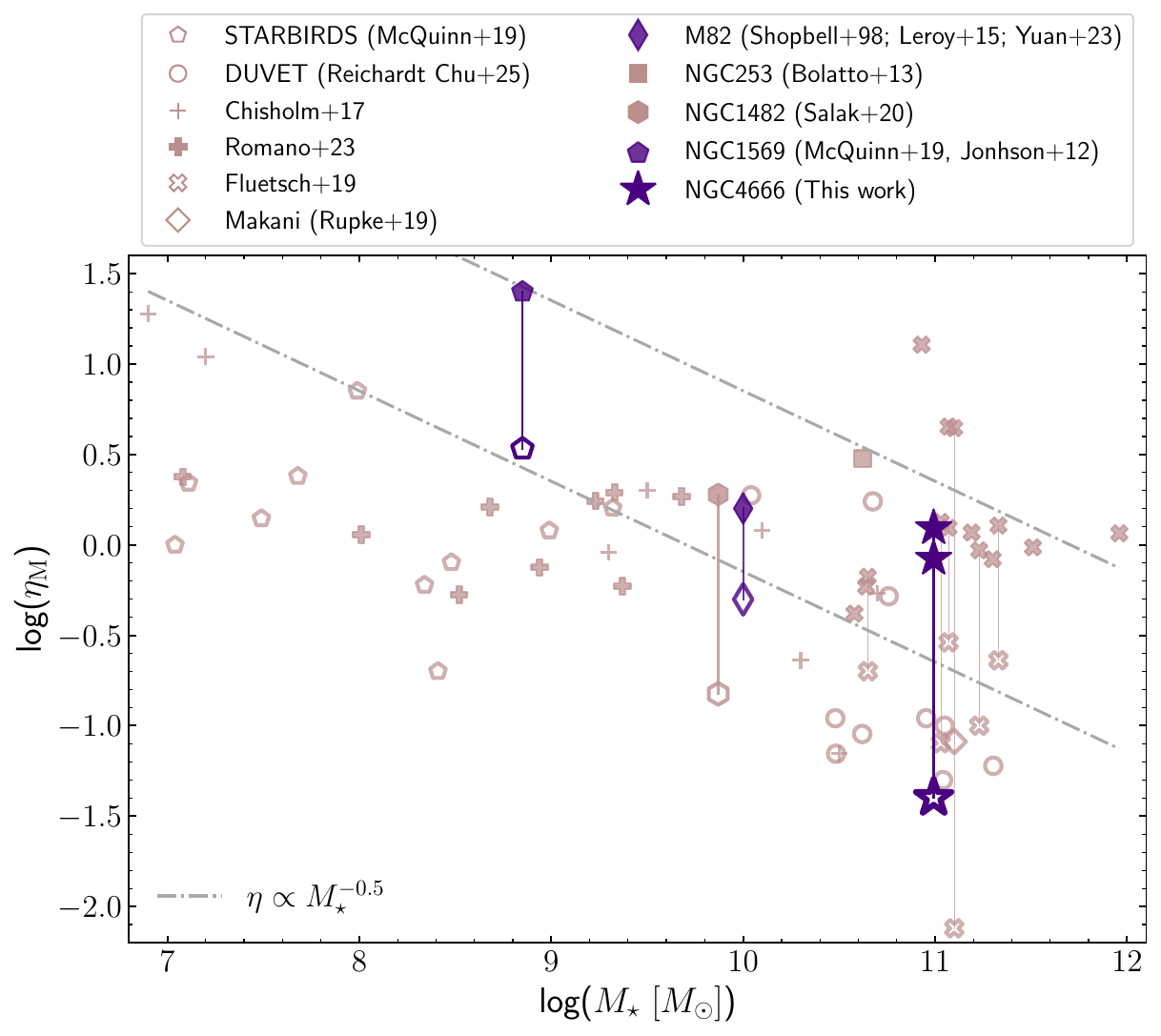}
    \caption{Mass loading factor measurements with respect to galaxy stellar mass for different samples in the Local Universe.
    Pale symbols represent previous works: empty symbols show ionised gas measurements, while filled symbols correspond to total mass loading factors combining ionised and molecular gas measurements (when available). Vertical lines connect measurements from the same galaxy.
    For M\,82, NGC\,1569 and NGC\,4666, all shown in indigo, the filled symbols represent mass loading factors summing ionised and neutral gas.
    The dashed grey lines represent theoretical scaling relations, with a slope similar to that of the FIRE-2 simulations. The lower line is normalised to pass through the median of the data, while the upper line is shifted by one order of magnitude.}
    \label{fig:comparison_simus}
\end{figure}

Figure~\ref{fig:comparison_simus} shows the relationship between mass loading factor and galaxy stellar mass, comparing NGC\,4666 to other galaxies in the local Universe.
The grey dashed lines represent two different $\eta_{\rm M}\propto M_{\star}^{-0.5}$ scaling relations with a normalization that differs by one order-of-magnitude. This scaling is consistent with the general trend of decreasing mass loading with increasing stellar mass found in many simulations \citep[e.g.][]{Muratov2015,Pandya2021}.
The ionised gas mass loading for NGC\,4666 is typical for its stellar mass when compared to other samples. Differences in mass loading from galaxies like M\,82 can be attributed to this trend with stellar mass.

Similar to CO observations \citep[e.g.][]{Fluetsch2021}, the colder phase gas carries significantly more mass in the outflow but at lower velocity than the ions, and is unlikely to reach the same distances into the circumgalactic medium (CGM) as the ionised gas.
Across M\,82, NGC\,1569 and NGC\,4666, resolved \ion{H}{I} observations consistently show mass loading factors 5-10 times larger than those of ionised gas, mirroring the molecular-to-ionised gas mass loading ratios found by \citet{Fluetsch2021}. These observations point to a picture in which the neutral phase (either molecular or atomic) exhibits mass loading factors an order of magnitude larger than for ionised gas. It is important to note, however, that this enhanced mass loading refers specifically to gas leaving the disk. The observed velocities of both \ion{H}{I} and CO indicate that this cold material does not travel far beyond $\sim$10~kpc in star formation-driven winds, confirming that while neutral gas dominates the mass budget of galactic outflows, it remains confined to the inner regions compared to the more extended ionised component.

\subsection{Energy loading factors of NGC\,4666 wind}
\label{sec:energy_loadings}
In this section, we consider the energy of the wind, where the energy rate is defined as $\dot{E} = 0.5 \dot{M} v^2$. To estimate the total energy released by supernovae, we assume that each core-collapse supernova (SN~II) explosion supplies $\rm 10^{51}ergs$ \citep{Murray2005} and that the ratio of supernovae rate to star formation rate is $0.01~{\rm SNe~II}/M_{\odot}$ \citep[for a Kroupa initial mass function,][]{Kroupa2001}. The star formation rate of the region underneath the wind is 3.7~$M_{\odot}~$yr$^{-1}$, which gives $\sim$0.04~SNe/year. Considering the total SFR of 10.5~$M_{\odot}~\rm yr^{-1}$ gives $\sim$0.11~SNe/year.
The energy released by star formation also depends on the efficiency of energy transfer from SNe to the surrounding environment. This efficiency is not well constrained because it depends on the properties of the interstellar medium. Models suggest that the efficiency of energy transfer from supernovae to the ISM can be as low as 10\% in the densest cores of nuclear starburst \citep[e.g.][]{Thornton1998} but more recent observations of galactic winds suggest greater values \citep[see details in][]{Veilleux2005}. For our calculation of $\dot{E}_{\rm SF}$, we consider an efficiency of 50\%. We find a kinetic energy supplied by supernovae of $\dot{E}_{\rm SF, total}=1.7\times10^{42}~\rm erg~s^{-1}$.
An efficiency of 10\% instead would lead to an injected energy by star formation five times lower, and thus to energy loading factors five times greater than our values.

\begin{figure}
    \centering
    \includegraphics[width=1\columnwidth]{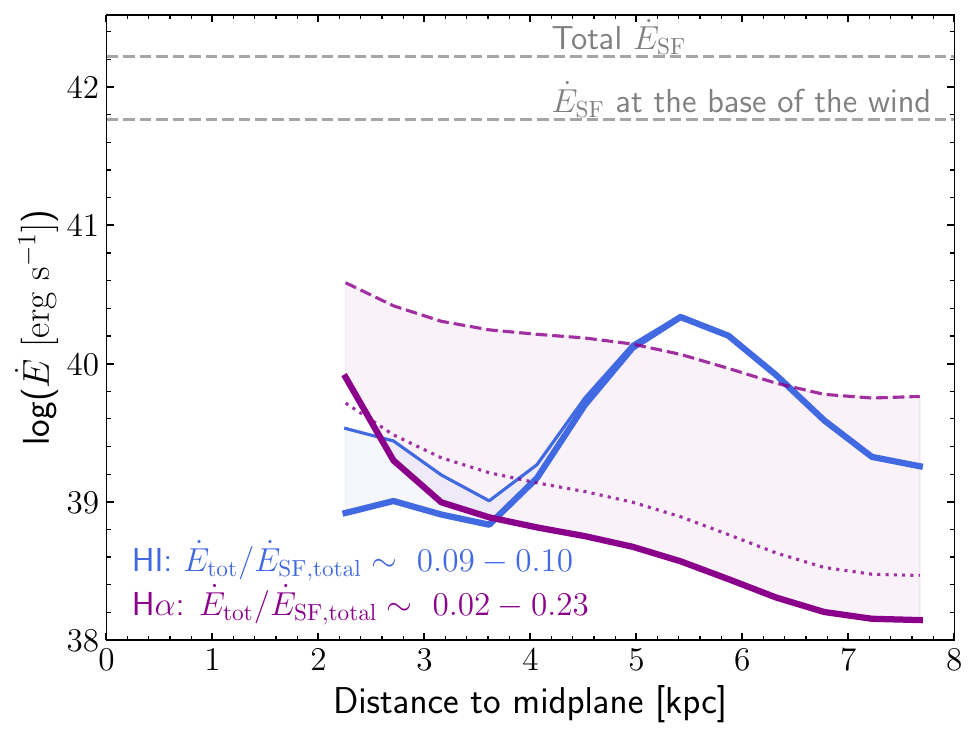}
    \caption{Energy outflow rates against distance to midplane. The blue curves show the \ion{H}{I} mass outflow rate measurements: the upper thin line corresponds to measurements before accounting for projection effects, while the thicker line presents corrected estimates (see Sect.\,\ref{sec:hi_mdotout}).
    The magenta curves display H$\alpha$ energy outflow rate measurements, assuming a clumping factor $C_e=1$. 
    The upper dashed line corresponds to the low-density assumption where $n_e$ decays as $z^{-0.8}$; the dotted line represents measurements for a constant $n_e=100~\rm cm^{-3}$; and the bottom solid line is the profile for the data-driven profile of $n_e$, given by Eq.\,\ref{eq:datadriven_ne}.
    The grey dashed lines indicate energy rates injected by supernovae feedback: the bottom line shows value for the region underneath the outflow, while the upper line represents the total energy rate injected from the whole galaxy, $\rm \sim 1.7\times10^{42}~erg~s^{-1}$.
    Values in the bottom-left corner give total energy loading factors for each phase, calculated by summing the measurements beyond 2~kpc.}
    \label{fig:edotout}
\end{figure}

Figure~\ref{fig:edotout} shows the radial profiles of energy outflow rate for both ionised and \ion{H}{I} gas phases.
For the atomic gas phase beyond 2~kpc, we measure $\dot{E}_{\rm out,\ion{H}{I}}=\rm 1.5\times10^{41}~erg~s^{-1}$ (after inclination/beam correction), corresponding to an energy loading factor ($\eta_{\rm E}=\dot{E}_{\rm out}/\dot{E}_{\rm SF}$) of $\sim$0.09 when considering the total star formation rate.
The ionised gas shows $\dot{E}_{\rm out,H\alpha}=\rm 2.9\times10^{40}~erg~s^{-1}$ assuming our data-driven electron density profile, yielding an ionised energy loading factor of 0.02.
If we calculate the kinetic energy injected from the centrally located starburst only, this increases these energy loading factors by approximately a factor of 2.8, giving a combined \ion{H}{I} + ionised energy loading of $\sim$0.35.

A complete summary of mass and energy loading factors for different phases and assumptions is given in Table~\ref{tab:outflow_measurements}.

We note that the energy rate profile of \ion{H}{I} presents sub-structure that is not observed in the H$\alpha$ profile. Specifically, the $\dot{E}_{\rm out, \ion{H}{I}}$ increases between 4 and 7~kpc. In contrast, $\dot{E}_{\rm out, H\alpha}$ decreases systematically across the entire range that we probe. The increase in $\dot{E}_{\rm out, \ion{H}{I}}$ is co-located with a rise in the velocity profile of the \ion{H}{I} gas shown in Fig.~\ref{fig:velocity_offsets}. Given that energy loading includes $v^3$, this is likely the cause of this bump. 

We note that the beam smearing could impact the velocity profile, which then may impact $\dot{E}_{out}$. The velocities of \ion{H}{I} are determined from the moment-1 decomposition to the \ion{H}{I} datacube. If the velocity near the galaxy includes a lower velocity component, from the disk, this would lower the measured moment-1 velocity. The large beam of the WALLABY data could extend that low velocity gas further into the disk, which may create lower velocities at $z\sim1-3$~kpc. It would have diminishing impact at larger velocity. The $v^3$ dependence of the outflow energy profile may increase this effect. Higher spatial resolution data will be helpful in these cases. 

Our combined \ion{H}{I} + ionised energy loading factor of $\sim$0.11 (with SFR$_{\rm total}$) or $\sim$0.35 (with SFR$_{\rm central}$) can be compared with other starburst-driven winds.

In M\,82, the combined energy rates from ionised and atomic gas account for  $\sim$10-12\% of the kinetic energy injected by the starburst \citep{Xu2023}. This is very similar to the energy loading we report for NGC~4666. 
 
We discuss below that we do not detect extraplanar CO. The frequency and amount of molecular gas in outflows is not understood \citep[see][]{Vijayan_Krumholz_2024}. We do not know if significant amounts of molecular gas would survive disk breakout in NGC\,4666. Moreover, the CO in M\,82 does not extend far beyond $\sim$2-3~kpc \citep{Krieger2019}, which is where we begin calculating our mass-outflow rate.

\renewcommand{\arraystretch}{1.3}
\begin{table*}
\caption{Total outflow measurements from different phases in NGC\,4666. The values for mass, mass rate, energy rate, mass loading, and energy loading are calculated based on gas emission beyond 2~kpc from the midplane on the upper side of the wind. These values are then multiplied by two to account for the contribution from the opposite side of the wind.
For CO emission, the derived values are limits based on the sensitivity of the observations (see Sect.~\ref{sec:co_mass}).}
\label{tab:outflow_measurements}
\centering
\adjustbox{width=\textwidth,center}{
\begin{tabular}{c|c|c|c|c|c|c|c|c|c}
\hline
Phase & Tracer & $v_{\text{out}}$ & Assumptions & $M_{\text{out}}$ & $\dot{M}_{\text{out}}$ & $\dot{E}_{\text{out}}$ & $\dot{M}_{\rm out}$/SFR$_{\rm total}$ & $\dot{E}_{\rm out}/\dot{E}_{\rm SF, total}$ & \\ 

 &  & [$\rm km~s^{-1}$] & & [$M_{\odot}$] & [$M_{\odot}\rm~yr^{-1}$] & [$\rm erg~s^{-1}$] &   \\ \hline

\multirow{3}{*}{Warm ionised} & \multirow{3}{*}{H$\alpha$} & \multirow{3}{*}{210 - 320} & \text{data-driven} $n_e$, $C_e = 1$ &  $4.4\times10^6$ & 0.5 & $2.9\times10^{40}$ & 0.04 & 0.05 \\  \cline{4-10} 
 
 & & & \text{decaying} $n_e$, $C_e = 1$ & $4.0\times10^7$ & 4.7 & $3.9\times10^{41}$ & 0.4 & 0.7 \\   \cline{4-10} 
 
 & & & \text{constant} $n_e$, $C_e = 1$ & $4.4\times10^6$ &  0.5 & $3.6\times10^{40}$ & 0.05 & 0.06 \\  \hline
 
\multirow{2}{*}{Atomic} & \multirow{2}{*}{\ion{H}{I}} & \multirow{2}{*}{150} &Inclination + beam corrected & $2.2\times10^8$ & 8.5 & $1.5\times10^{41}$ & 0.8 & 0.09 \\
    & & & -- & $4.1\times10^8$ & 13 & $1.7\times10^{41}$ & 1.2 & 0.10 \\ \hline
Cold molecular & CO(1-0) & & $v_{\rm CO}=v_{\rm \ion{H}{I}}$ & $\leq4\times10^{7}$ & $\leq2.9$ & $\leq2\times10^{40}$ & $\leq0.3$ & $\leq0.04$ \\ 
\hline
\end{tabular}
}
\end{table*}

\subsection{Molecular gas in the wind}
\label{sec:co_mass}
\begin{figure}
    \centering
    \includegraphics[width=1\columnwidth]{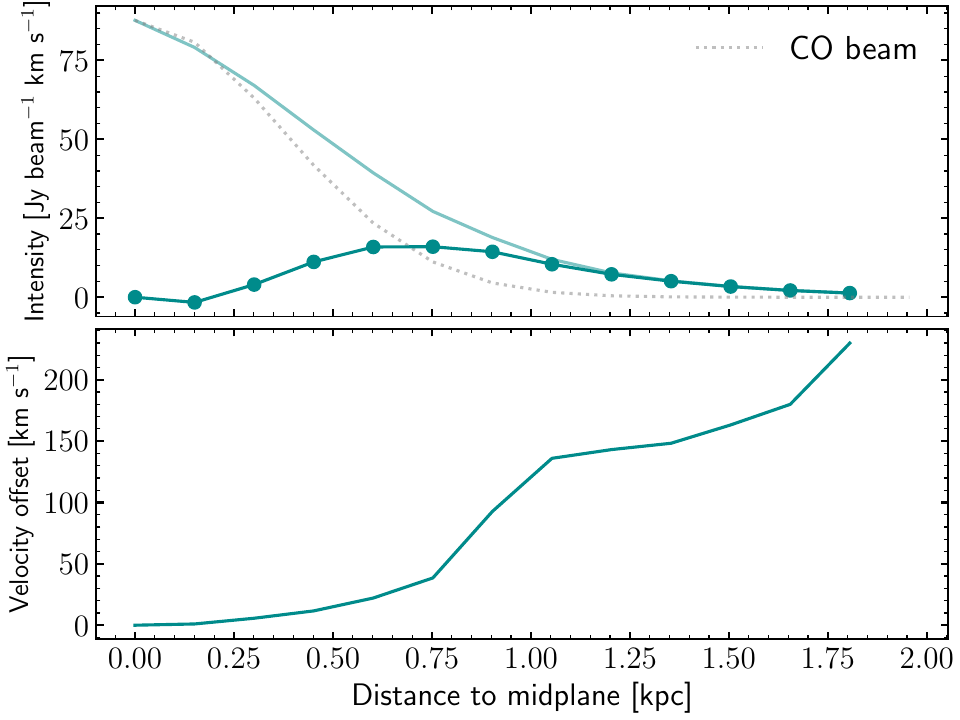}
    \caption{Minor-axis profiles of CO emission. The top panel represents the velocity-integrated flux intensity, computed from the $5\sigma$-clipped moment 0 map. The dashed grey line shows the beam profile, the transparent solid line shows the CO flux profile, and the darker solid line with points shows the flux profile after subtracting the beam profile.}
    \label{fig:co_profiles}
\end{figure}

At a projected distance of approximately 1.75~kpc, we do not detect any extraplanar molecular gas in NGC\,4666. The top panel of Fig.~\ref{fig:co_profiles} shows the CO(1–0) velocity-integrated intensity as a function of distance from the midplane.
The rms noise level per channel in the CO(1-0) cube is 11.1~mJy~beam$^{-1}$, with a a channel width of 20~km~s$^{-1}$ \citep{Lee2022}. Assuming a line width of 200~km~s$^{-1}$, motivated by the integrated spectrum between 1 and 2~kpc, we can compute a minimal detectable CO mass per beam in the galaxy.
A circular synthesised beam covers an area of $\rm \sim1.2~kpc^2$. Based on this CO mass limit and the area of the upper-side wind, we estimate an upper limit for the molecular gas mass in the outflow. The derived CO mass limit is $1.9\times10^7~M_{\odot}$, assuming a CO-to-H$_2$ conversion factor of $\alpha_{\rm CO}=1~M_{\odot}~\rm(K~km~s^{-1}~pc^{-2})^{-1}$ , consistent with values adopted for other starburst-driven winds \citep[e.g.,][]{Bolatto2013b, Leroy2015}.
To estimate the total molecular gas outflow rate, we multiply this mass limit by 2 to account for the lower side of the wind, and assume the molecular gas has the same velocity as the atomic gas ($v=150~\rm km~s^{-1}$). 
This gives an outflow mass rate limit of $\dot{M}_{\rm out, CO, lim}=~2.9~M_{\odot}~\rm yr^{-1}$, and a limit of $\dot{M}_{\rm out,CO,lim}/\rm SFR_{total}$=0.3, which is significantly lower than the mass loading of \ion{H}{I} and falls within the range of the ionised mass loading (0.04-0.4). This limit is much lower than measured in nearby starburst like NGC~253 \citep{Krieger2019} and M~82 \citep{Leroy2015}.
Observations of M\,82 indicate an atomic-to-molecular mass rate ratio of about 0.7 at 2.2~kpc. Assuming a similar atomic-to-molecular ratio for NGC\,4666 would imply a  molecular outflow mass rate of $\sim12~M_{\odot}\rm~yr^{-1}$, comparable to the SFR of NGC\,4666. This is much larger than our limit.

\section{Properties of the H$\alpha$ bubble}
\label{sec:bubble}
On the far right side of Fig.~\ref{fig:ha_bicone} we identify a secondary region extending outward from the disk that sits over a small-scale enhancement in the $\Sigma_{\rm SFR}$. The region appears to be distinct from the biconical wind in the centre of the galaxy. \cite{Schneider2024} found in simulations that the spatial distribution of the star formation in a galaxy may have an impact on the winds. We, therefore, consider the properties of this ``bubble''. 

We find that the bubble extends to a distance of $\sim$3.7~kpc from the midplane, after which the H$\alpha$ surface brightness is consistent with the noise in the narrow-band image. Using the H$\alpha$ emission, we determine a width of the bubble along the major axis to be $\sim$2.5~kpc, and it is located $\sim$7~kpc from the galaxy centre. \cite{Barnes2023} study voids in the face-on spiral galaxy NGC\,628. They interpret these voids as the result of star formation feedback. They find that the ionised gas is brightest in the shells of such structures. If the bubble we observe is a similar phenomenon that is viewed edge-on, then its diameter is larger than the biggest void not only in NGC\,628, but in the entire PHANGS sample \citep{Watkins2023}. NGC\,4666 has a stronger starburst than those galaxies, and therefore it makes sense that feedback drives larger structures. That the bubble extends to higher latitudes above the disk than its width is consistent with standard expectations for superbubbles \citep[e.g.][]{MacLow1988}. 

While there is significant \ion{H}{I} above the disk in this region it is not clear that it is an enhancement over the surrounding emission, as is seen in H$\alpha$. Although the neighboring galaxy NGC\,4668 is located on the opposite side of the disk from the region where the bubble is observed, we cannot rule out the possibility that the \ion{H}{I} at the edge of the disk traces an interaction with this nearby companion.
Higher spatial resolution observations may be required to study \ion{H}{I} in these structures. We will limit our analysis to the ionised gas, in which the morphology clearly resembles a filamentary structure that connects to the disk. 

\begin{figure*}
    \centering
    \includegraphics[width=.98\textwidth]{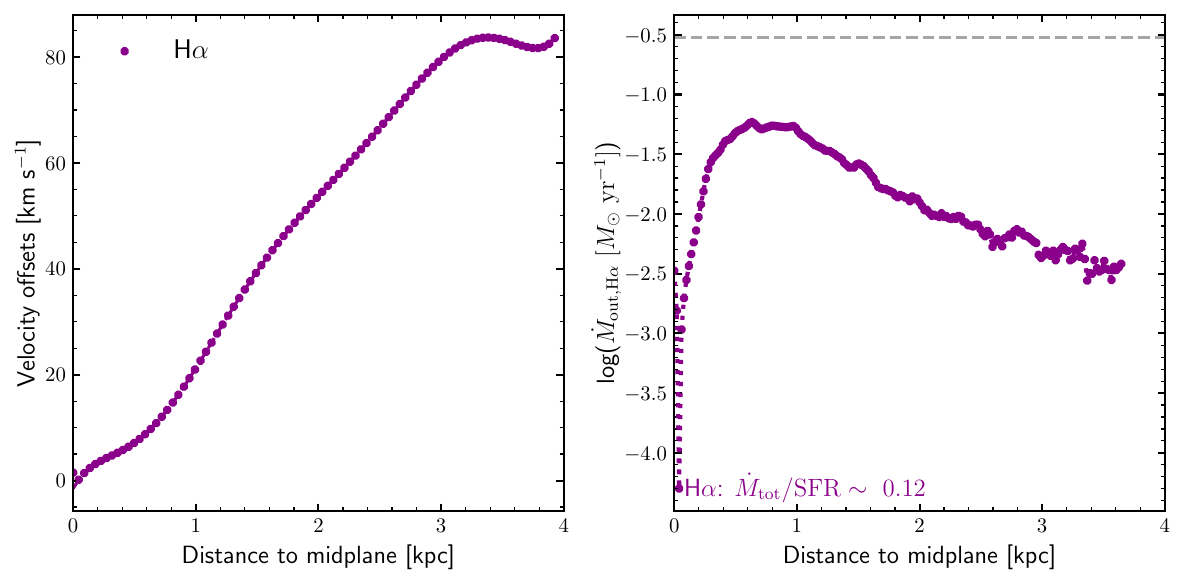}
    \caption{Ionised gas velocity and mass outflow rate profile of the bubble.
    The velocity H$\alpha$ measurements are based on the GECKOS MUSE observations, which coverage stops at $\sim$4~kpc from the midplane. The horizontal dashed grey line in the right panel represents the 24$\micron$-based SFR value of the region underneath the extraplanar emission. The outflow rate is computed assuming a constant $n_e=60~\rm cm^{-3}$, which is what we measure in the galactic disk below the bubble.}
    \label{fig:vel_offset_bubble}
\end{figure*}

The velocity offset of H$\alpha$ gas is $\sim$80~km~s$^{-1}$ at the detected edge of the bubble (3.7~kpc), as shown in the left panel of Fig.~\ref{fig:vel_offset_bubble}. This translates to a travel time of $t_{\rm dyn}\sim z/v\sim 45$~Myr. The SFR in the disk under the bubble is $\sim$0.3~M$_{\odot}$~yr$^{-1}$. If we treat this feature as an energy-driven bubble \citep[following][]{ThompsonHeckman2024}, we derive an expansion timescale of $\sim$10~Myr, which given the systematic uncertainties in the calculation is in good agreement.

Using the MUSE observations, we find that the ionised gas in the bubble region is slower than the gas within the biconical wind over the galaxy centre. At 4~kpc above the midplane, the ionised gas velocity is $\sim$80~km~s$^{-1}$ in the bubble. At the same distance above the midplane the central outflow has a velocity of $\sim$170-280~km~s$^{-1}$. The ratio of these velocities is consistent with simple expectations of energy-driven winds \citep{Chen2010,ThompsonHeckman2024}, and with observed scaling relations of resolved outflow velocity with $\Sigma_{\rm SFR}$ \citep{ReichardtChu2022a,ReichardtChu2024}.

There is insufficient signal-to-noise on the [\ion{S}{II}] doublet to estimate the electron density beyond the galaxy midplane. In the galaxy midplane, below the bubble, we measure $n_e\sim57^{+15}_{-20}$~cm$^{-3}$. It is not possible to robustly determine the electron density at any higher latitude.

We estimate a mass outflow rate of the ionised gas for the bubble using a similar calculation as before. We make a simple assumption that the electron density of the bubble gas is similar to the disk below it, $n_e\sim 60$~cm$^{-3}$. This yields $\dot{M}_{\rm out}\sim5\times10^{-3}$~M$_{\odot}$~yr$^{-1}$, which translates to a mass-loading of $\dot{M}_{\rm out}/\rm SFR \sim 0.02$, for all gas above $2$~kpc from the midplane. The right panel of Fig.~\ref{fig:vel_offset_bubble} shows the resulting outflow mass rate profile.

In Fig.~\ref{fig:tdepl}, we show that the region of the disk that is underneath the bubble shows an increase in $\Sigma_{\rm SFR}$ from surrounding regions, and it is higher than the opposite side of the disk, which has no visible bubble. This region has a peak $\Sigma_{\rm SFR}$ that is less than half that of the centre of the galaxy. If we assume that the bubble is launched from a region that is symmetric within the disk we can estimate the associated SFR surface density as $\Sigma_{\rm SFR}={\rm SFR}/\pi R^2 \sim 0.08$~M$_{\odot}$~yr$^{-1}$~kpc$^{-2}$, where $\rm SFR$ and $R$ are the SFR and half-width of the bubble. The mass loading of the bubble is significantly lower than that found in the DUVET sample of resolved outflows in face-on galaxies. \cite{ReichardtChu2024} find for $\Sigma_{\rm SFR}\sim 0.1-0.01$~M$_{\odot}$~yr$^{-1}$~kpc$^{-2}$ that mass-loading ranges from 0.1-1. We have discussed already that edge-on galaxies may be biased towards lower $\dot{M}_{\rm out}$, due to observational effects. For example, if we instead measure all gas above $\sim$1~kpc this yields a much higher $\dot{M}_{\rm out}\sim0.04$~M$_{\odot}$~yr$^{-1}$, which translates to $\dot{M}_{\rm out}/\rm SFR \sim 0.1$. This would be within the range estimated by \cite{ReichardtChu2024}. Both observations of face-on and edge-on outflows have their own advantages: face-on orientations enable more robust velocity measurements, while edge-on views better constrain the vertical extent and distance travelled by the ejected material. Comparison of these different measurements is therefore highly valuable. Clearly, more work comparing outflow measurements in face-on and edge-on galaxies would be helpful to place these, and future, measurements of outflows in context.

\section{Summary and conclusions}
\label{sec:summary}
In this paper we use observations from VLT/MUSE, narrow-band H$\alpha$ imaging, ASKAP \ion{H}{I}, and ALMA/ACA CO to study the galactic-scale wind of the nearby, starbursting galaxy NGC\,4666. Our observations are the first measurements of mass outflow rates in multiple gas phases in this galaxy. In Fig.~\ref{fig:ha_bicone}, we determine the geometry of the biconical outflow to have an opening angle of $\sim25-30\degr$ and the edge of the H$\alpha$ bifrustrumal cone extends to $\rm \sim8~kpc$ from the galaxy midplane. Based on \ion{H}{I} and H$\alpha$ emission, we measure a multiphase mass outflow rate of $5-13~M_{\odot}~\rm yr^{-1}$.
The wind is dominated by the atomic phase traced by \ion{H}{I}, with a mass outflow rate more than an order-of-magnitude higher than that of the warm ionised phase traced by H$\alpha$.

\subsection{Implications of high electron density in the wind of NGC\,4666}
Our measured mass outflow rate in NGC\,4666 is impacted by our observations that the electron density profile of NGC\,4666 wind gas does not systematically decay (Fig.~\ref{fig:ne_niiha_profiles}). We observe that the electron density first declines up to a scale-height of $\sim$2~kpc, then rises again to values of $n_e\sim100-300$~cm$^{-3}$. The observed electron density remains constant from $z\sim$ 3~kpc, until we run out of sufficient signal-to-noise to estimate $n_e$. As discussed in the text, the electron density is a difficult quantity to measure, and requires significant signal-to-noise. While this discovery is enabled by the deep GECKOS observations, one still must take care to identify the possible systematic uncertainties. For example, the fraction of gas that has sufficient [\ion{S}{II}] flux to measure the electron density decreases rapidly as a function of distance from the galaxy. Therefore, the high electron densities at large distance from the midplane are only securely measured on the brightest [\ion{S}{II}] gas, which may be biased to elevated shock-heating \citep[e.g.][]{Dopita1995}. Additionally, ionised gas emission lines are more sensitive to dense gas, potentially biasing measurements toward the higher-density gas if dense clouds are embedded in a lower-density medium.
Observations that decompose the spectra of face-on outflows into broad and narrow components have found similar values of the outflow electron density \citep[e.g.][]{ForsterSchreiber2019,Fluetsch2021}. While those observations do not have spatial information, they do imply that high values are not unprecedented. Fisher et al. {\em submitted} extend this result with spatially resolved measurements of $n_e$ in six GECKOS galaxies and find that rising density profiles are a common feature of strong outflows. Verna et al. {\em in prep} will likewise show a similar result, indicating high $n_e$ in the outflow with gas located further than 2~kpc from the midplane in M\,82.

A variety of recent high-resolution simulations of starburst-driven outflows predict a decay in the density profile in outflows \citep{Girichidis2016,Kim2018,Schneider2020}. Similarly, monotonically decaying profiles are seen in both classic \citep{ChevalierClegg1985} and modern \citep{FieldingBryan2022} theories describing outflows. The observations shown in Fig.~\ref{fig:ne_niiha_profiles}, therefore, suggest that some physical process is missing from those simulations. An interaction of outflow gas with a highly pressurised circumgalactic medium could plausibly alter the density of gas. Those simulations do not, typically, include this. Alternatively, thermal instabilities within the wind itself may lead to changes to the density profile \citep{Thompson2016,Nguyen2024}.   

The electron density profile of NGC\,4666 also has implications for estimation of the mass profile in outflows. For ionised gas, the mass is inversely proportionate to the electron density. The rise in $n_e$ at high latitude then implies that the ionised gas mass is more concentrated toward the galaxy than one would assume with a decaying profile of $n_e\propto z^{-1}$. Indeed, we measure an order-of-magnitude increase in $\dot{M}_{\rm out}$ for the decaying profile of $n_e$ compared to that motivated by our observations. If this is true, then this ought to be reproduced in simulations of winds. Moreover, we must take this into account when comparing with observations that do not have empirical measurements of $n_e$ that extend beyond $\sim$2~kpc.

The clumpiness of the ionised gas remains a significant source of uncertainty in the mass calculations. If a physical process is creating the increase in $n_e$ it is plausible to expect this to likewise generate a change in the clumpiness ($\langle n_e^2 \rangle/\langle n_e\rangle^2$) of that same gas. \citet{Yuan2023} highlight that uncertainties in this factor can strongly affect derived outflow masses. If, for example, $\langle n_e^2 \rangle/\langle n_e\rangle^2$ increases by a factor of 10$\times$ over the same range where $n_e$ rises, it would cancel out the impact on the difference in masses. Recent observations with both \textit{JWST} \citep{Fisher2025} and \textit{HST} \citep{Lopez2025arXiv} demonstrate that it is now possible to resolve cloud-scale properties of outflows. More observations, beyond M\,82, are direly needed to finally estimate accurate masses of galactic winds. 

\subsection{\ion{H}{I} and ionised outflow mass loading}
For gas beyond 2~kpc in NGC\,4666, we find that the neutral gas dominates the mass loading in the wind. Our measurements yield $\dot{M}_{\rm out, ion}/\dot{M}_{\rm out, \ion{H}{I}}\sim1/17$ after correcting the \ion{H}{I} measurements for inclination and beam effects, and assuming our observed $n_e$ profile.
This ratio is sensitive to the assumed electron density profile and systematic uncertainties in the \ion{H}{I} measurements. 
The ratio increases to $\sim$1/1.8 assuming that the electron density decays as $z^{-1}$, as is often done in the literature.
Adopting a more conservative approach for the \ion{H}{I} measurements to exclude gas beyond 5~kpc from potential group interaction contamination (while maintaining our observed $n_e$ profile) gives $\sim$1/11.
Even with the most conservative approach, the neutral hydrogen mass outflow rate remains higher than that of the ionised gas.

Very few galaxies have mass outflow rates measured in both \ion{H}{I} and H$\alpha$, making comparisons challenging.
In M\,82, \citet{Xu2023} measure $\dot{M}_{\rm out, ion}/\dot{M}_{\rm out, \ion{H}{I}} \sim 1-2$, within $\sim$2~kpc from the disk. Nonetheless, this is not directly comparable to our measurement, as the inclination and resolution in NGC\,4666 prevent us from distinguishing disk and outflow light.

We find that the velocity of both ionised and \ion{H}{I} gas rises sharply in the outflow. The velocity of the ionised gas rises to $\sim$200-300~km~s$^{-1}$, consistent with previous observations of starburst-driven winds. Although the velocity of the \ion{H}{I} gas is roughly 50-75\% that of the ionised gas, both profiles show similar substructures, with rises occurring at comparable distances. Neither the \ion{H}{I} nor the H$\alpha$ gas has sufficient velocity to escape a Milky Way-mass galaxy halo ($\sim$500-600~km~s$^{-1}$). Moreover, the fact that the wind mass loading is dominated by the \ion{H}{I} gas suggests that the majority of mass leaving the disk in the outflow will not travel far beyond several kiloparsecs in the halo. This aligns with results from \citet{Marasco2023}, who measure the mass of gas in a sample of outflows that has $v_{\rm out}>v_{\rm escape}$ and find this to be far lower than predicted by large cosmological simulations.
Similar conclusions have been drawn from other observations \citep{Chisholm2015,Davies2019,ForsterSchreiber2019}.
Resolved observations like those presented here for NGC\,4666 provide unique constraints on galaxy simulations, which currently predict a wide range of outflow behaviours. Such observations are crucial for discriminating between feedback models that can all reproduce key galaxy scaling relations, including the stellar mass function and the star formation rate–stellar mass relation \citep[e.g.,][]{Pillepich2018}.

Many questions remain open about the cold gas in outflows. We do not know whether the electron density and \ion{H}{I} mass loading we observe are typical for galactic winds, or if NGC\,4666 is unique. Given that galactic winds are a critical and necessary component of essentially all modern galaxy evolution models \citep{NaabOstriker2017}, understanding the full mass loss from winds in all gas phases is essential. GECKOS is well positioned to explore the remaining questions. While progress on the ionised gas observations will continue, observations of \ion{H}{I} are necessary. NGC\,4666 is one of the nearest GECKOS targets, and therefore, among the only ones in which WALLABY resolution is sufficient. The ongoing upgrades aimed at establishing the early phases of the Square Kilometre Array will be critical to understanding how much mass is exiting galaxies in winds, and how far it travels. These are fundamental questions that require answers to build complete models of galaxy evolution.

\section*{Acknowledgements}

Writing of this draft was significantly improved through conversations with Drummond Fielding, Max Gronke, Resherle Verna and John Chisholm. We are grateful to Sarah Busch for ongoing software support. 

Parts of this research were supported by the Australian Research
Council Centre of Excellence for All Sky Astrophysics in 3 Dimensions (ASTRO 3D), through project number CE170100013.
GECKOS is based on observations collected at the European Organisation for Astronomical Research in the Southern Hemisphere under ESO program ID 110.24AS. We wish to thank the ESO staff, and in particular the staff at Paranal Observatory, for carrying out the GECKOS observations.
This paper makes use of services that have been provided by AAO Data Central (datacentral.org.au)
This paper makes use of the following ALMA data: ADS/JAO.ALMA\#2019.1.01804.S. ALMA is a partnership of ESO (representing its member states), NSF (USA) and NINS (Japan), together with NRC (Canada), NSTC and ASIAA (Taiwan), and KASI (Republic of Korea), in cooperation with the Republic of Chile. The Joint ALMA Observatory is operated by ESO, AUI/NRAO and NAOJ.
This scientific work uses data obtained from Inyarrimanha Ilgari Bundara / the Murchison Radio-astronomy Observatory. We acknowledge the Wajarri Yamaji People as the Traditional Owners and native title holders of the Observatory site. CSIRO’s ASKAP radio telescope is part of the Australia Telescope National Facility (https://ror.org/05qajvd42). Operation of ASKAP is funded by the Australian Government with support from the National Collaborative Research Infrastructure Strategy. ASKAP uses the resources of the Pawsey Supercomputing Research Centre. Establishment of ASKAP, Inyarrimanha Ilgari Bundara, the CSIRO Murchison Radio-astronomy Observatory and the Pawsey Supercomputing Research Centre are initiatives of the Australian Government, with support from the Government of Western Australia and the Science and Industry Endowment Fund.
This research has made use of the NASA/IPAC Infrared Science Archive, which is funded by the National Aeronautics and Space Administration and operated by the California Institute of Technology.

This work was supported by STFC grant ST/X001075/1.

MM acknowledges support from the UK Science and Technology Facilities Council through grant ST/Y002490/1.

LC acknowledges support from the Australian Research Council Discovery Project funding scheme (DP210100337).

FP acknowledges support from the Horizon Europe research and innovation programme under the Maria Skłodowska-Curie grant “TraNSLate” No 101108180, and from the Agencia Estatal de Investigación del Ministerio de Ciencia e Innovación (MCIN/AEI/10.13039/501100011033) under grant (PID2021-128131NB-I00) and the European Regional Development Fund (ERDF) ``A way of making Europe''.

THP gratefully acknowledges support from the National Agency for Research and Development (ANID) in form of the CATA-Basal FB210003 grant.

LASL is supported by Coordenação de Aperfeiçoamento de Pessoal de Nível Superior - Brasil (CAPES) - Finance Code 88887.637633/2021-0.

LMV acknowledges support by the German Academic
Scholarship Foundation (Studienstiftung des deutschen Volkes) and the Marianne-Plehn-Program of the Elite Network of Bavaria.

\section*{Data Availability}
The GECKOS survey is still in progress, but the MUSE NGC\,4666 data used in this work is available in the archive (program IDs: 096.D-0296(A) and 110.24AS.004).
Both the ALMA/ACA (project code: 2019.1.01804.S) and the ASKAP observations are publicly available.



\bibliographystyle{mnras}
\bibliography{ngc4666} 




\appendix

\section{Ionised gas maps of NGC\,4666 and its wind filament}
\label{app:gas_maps}
\begin{figure*}
    \includegraphics[width=0.9\textwidth]{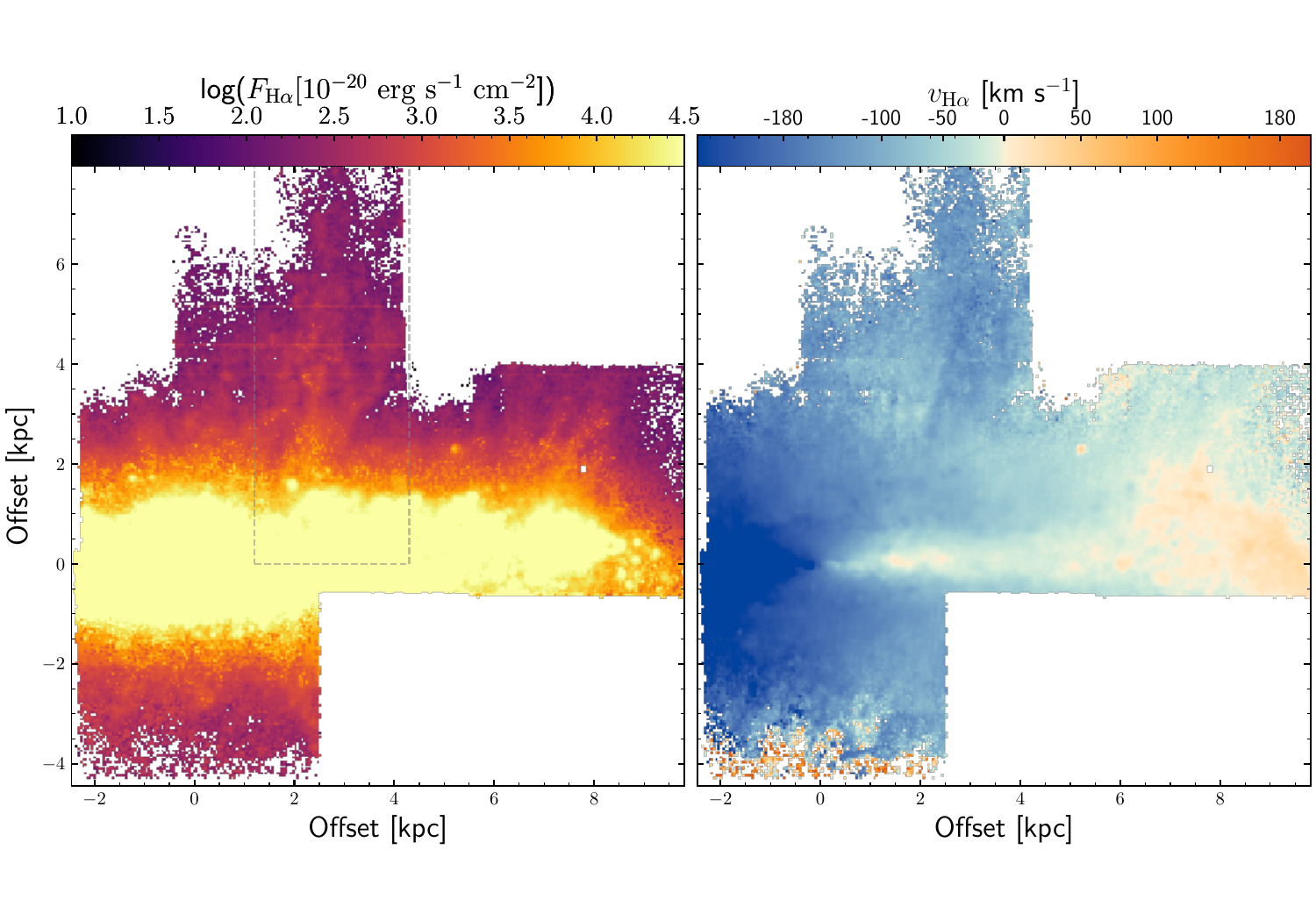}
    \caption{H$\alpha$ flux and velocity maps of NGC\,4666 from GECKOS VLT/MUSE observations. H$\alpha$ emission extends up to $\rm \sim~8kpc$, as well as extraplanar emission in the upper right region above the disk. The H$\alpha$ velocity shows rotation pattern but also some extraplanar gas that does not have the same kinematics as the disk underneath it. The upper-right filament region is depicted with a dotted grey rectangle and its corresponding gas property maps are shown in Fig.\,\ref{fig:filament_maps}.}
    \label{fig:ha_flux_vel_maps}
\end{figure*}

\begin{figure*}
    \includegraphics[width=\textwidth]{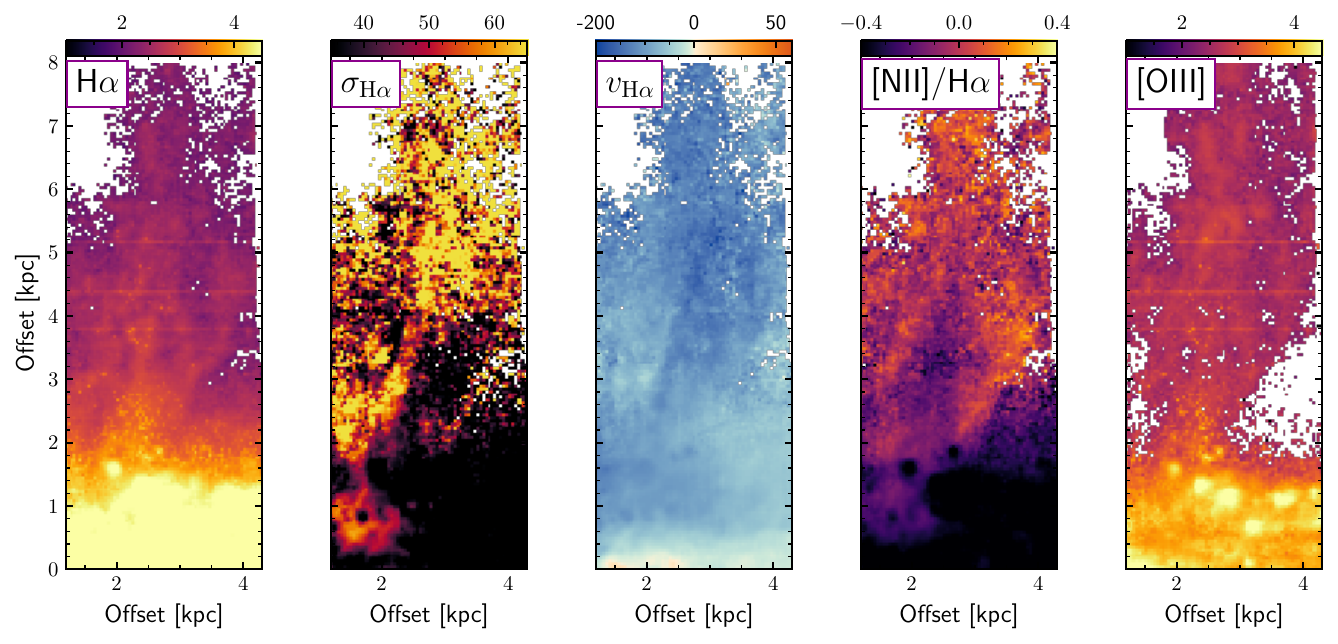}
    \caption{Ionised gas properties of the upper-right wind filament (grey dashed box in left panel of Fig.~\ref{fig:ha_flux_vel_maps}) derived from GECKOS VLT/MUSE observations. The first and last panel display the H$\alpha$ and [\ion{O}{III}] flux, in log($\rm 10^{-20}~erg~s^{-1}~cm^{-2}$), respectively. The second panel shows the H$\alpha$ velocity dispersion in $\rm km~s^{-1}$, corrected for instrumental dispersion. The third panel presents the H$\alpha$ observed velocity in $\rm km~s^{-1}$. The fourth panel corresponds to the flux ratio log([\ion{N}{II}]/H$\alpha$).}
    \label{fig:filament_maps}
\end{figure*}

In this appendix we show the spatially-resolved properties of the ionised gas, derived from our GECKOS observations.
Figure~\ref{fig:ha_flux_vel_maps} displays two different mosaicked maps (H$\alpha$ flux and velocity) for the entire MUSE coverage of NGC\,4666 and Fig.~\ref{fig:filament_maps} presents zoomed-in maps on the upper-right wind filament.

\section{HI column density profiles: minor axis vs. reprojected major axis}
\label{app:inclination_effect}
\begin{figure}
    \includegraphics[width=0.49\textwidth]{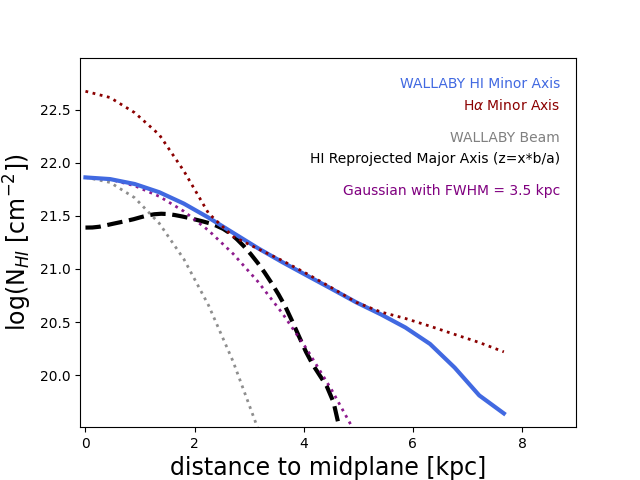}
    \caption{Comparison of \ion{H}{I} column density profiles as a function of projected distance perpendicular to the galactic midplane.
    The solid blue line shows the \ion{H}{I} column density profile extracted along the minor axis (perpendicular to the disk) from the WALLABY moment 0 map. The dashed black line shows the reprojected major axis profile, i.e., the radial distribution along the disk plane corrected for inclination. The dotted grey line indicates the beam profile, illustrating the instrumental resolution. For comparison, the dotted red line shows the H$\alpha$ flux profile extracted along the minor axis.
    The dotted purple line represents a Gaussian beam profile with FWHM=3.5~kpc that matches the reprojected major axis profile.}
    \label{fig:hi_column_profiles}
\end{figure}

This section compares \ion{H}{I} column density profiles extracted along different axes to assess the systematic effects introduced by galaxy inclination on the extraplanar \ion{H}{I} gas properties.
Figure~\ref{fig:hi_column_profiles} shows two \ion{H}{I} profiles: the \ion{H}{I} column density profile along the minor axis (solid blue line), i.e., perpendicular to the galactic disk, derived from the moment~0 map; and the reprojected major axis profile (dashed black line), which shows the radial distribution along the disk plane corrected for inclination effects using the major-to-minor axis ratio \( b/a = \cos(i) \) where \( i \) is the inclination angle of NGC\,4666. The WALLABY the beam profile (dotted grey line) illustrates the instrumental resolution limit.
The comparison reveals that, from about 3~kpc above the disk, the column density drops slower in the minor axis profile than in the major axis profile.
For reference, the dotted red line shows the H$\alpha$ profile extracted along the minor axis. By contrast, the minor axis \ion{H}{I} profile and H$\alpha$ profile show similar slopes over the range $z\sim2−6$~kpc, suggesting that both the neutral and ionised gas phases experience similar physical processes.

We note, however, that this model assumes the \ion{H}{I} disk has zero intrinsic thickness, which is not the case in reality. Accounting for a finite thickness may reduce the mass loading by a small amount, though that thickness is likely marginal compared to the larger distances we consider.


\bsp	
\label{lastpage}
\end{document}